# Corner cutting connects chiral colorimetry to net electric flux in lossless all-dielectric metasurfaces

Zaid Haddadin[1], Anna My Nguyen[2], and Lisa V. Poulikakos[2,3,*]

[1]*Department of Electrical & Computer Engineering, UC San Diego, La Jolla, CA 92093-0021, USA*
[2]*Department of Mechanical & Aerospace Engineering, UC San Diego, La Jolla, CA 92093-0021, USA*
[3]*Program of Materials Science & Engineering, UC San Diego, La Jolla, CA 92093-0021, USA*
*[\*lpoulikakos@ucsd.edu](mailto:lpoulikakos@ucsd.edu)*

**Abstract:** All-dielectric metasurfaces can produce structural colors, but the most advantageous design criteria are still being investigated. This work numerically studies how the two-dimensional shape of nanoparticles affects the colorimetric response under circularly polarized light (CPL) to develop a sensor distinguishing CPL orientations. Using lossless dielectric materials (silicon nitride on silicon dioxide), we achieve far-field dichroism by modifying oblong nanoparticles into L-shaped structures through corner cuts. This design suppresses one electric dipole under CPL illumination, leading to differential colorimetric responses. We link these responses to a decoupling effect in the near-field net electric flux. Our findings provide design guidelines for all-dielectric, lossless colorimetric sensors of chiral light.

## 1. Introduction

Circularly polarized light (CPL) can be classified by its handedness: clockwise or counterclockwise [1]. This classification depends on the rotation direction of a propagating electric field vector [1,2]. Tracing the paths of the electric field vectors in the clockwise and counterclockwise orientations will reveal distinct helical structures that are not superimposable with each other through rotation or translation operations, despite being mirror images of one another [2]. This lack of reflection symmetry underpins the chiral phenomenon [3].

Chirality is an inherent structural property, observed across scales from the subatomic to the galactic [4]. It is a fundamental feature of life on Earth [5]. The widespread presence and essential role of chirality in the life sciences underscore the necessity of recognizing and distinguishing between handed structures [5,6]. One consequence of failing to do so is illustrated in the limb malformations caused by the thalidomide disaster of the late 1950s [7,8].

This work does not focus on pharmaceuticals or biological structures. Instead, it addresses the challenge of distinguishing between the clockwise and counterclockwise handed orientations of incoming CPL. This task has been used to describe the structure and distribution of collagen fibrils [9], characterize the structure and conformational changes of proteins [10], and track rotational movement in bacterial species [11]. The techniques used in these studies fall under the umbrella of circular dichroism (CD) spectroscopy [10,12].

CD spectroscopy relies on the differential extinction between the clockwise and counterclockwise orientations of CPL by chiral matter [12]. However, detecting CD signals in biological matter is challenging due to their weak nature [13], prompting research into methods for enhancing these signals [2,13–16]. Here, we examine how a class of artificial materials – metasurfaces [17] – can be designed to provide differential responses to clockwise and counterclockwise CPL [18–20].

Metasurfaces are composed of many constituent sub-wavelength elements arranged along a two-dimensional surface [21]. The material choice and design of these elements affect the physical properties of the metasurface [17,21]. Some designs can make them sensitive to the incoming polarization state of light [22]. For example, using chiral geometries for the constituent elements can result in differential responses between clockwise and counterclockwise CPL [23] – the focus of this work.

We limit our investigation to the visible light regime [24], which influences our choice of material. Metasurfaces can be made of plasmonic and/or dielectric materials [25]. Plasmonic systems suffer from high optical losses in the visible light regime [26–28]. In contrast, dielectric materials can exhibit low dissipative losses [29]. Among these, silicon nitride shows no losses in the visible light regime [30], allowing the entire incoming CPL signal to be translated into a far-field measurement [31,32]. Because this work is limited to the visible light regime, it is possible to predict a color viewable by the human eye from the far-field spectroscopic measurements [33,34]. Silicon nitride is known to be an efficient generator of structural color when metasurfaces are designed to support Mie resonances, as this work does [35].

However, our interest extends beyond observations of the far-field spectra. Our work asks, "why do these observations occur?" To answer this, we examine the near-field effects within and around the metasurface elements. We analyze the electric field enhancement values and the net electric flux. Our goal is to build correlations between the geometric designs of the constituent elements, the consequence of the design choices on the near field, and the resultant far-field spectra. We hypothesize that certain geometric parameters can induce a decoupling in the propagation of the in-plane electric field vectors under illumination of one handedness of CPL but not the other; and that these decoupled vectors correlate with differential far-field spectra. By translating the differential far-field spectra into a color, we realize our correlations open a path to possibly achieving maximal colorimetry differentiation between clockwise and counterclockwise CPL illumination. Thus, the potential for chiral colorimetry in Mie resonator arrays can be found in net electric flux observations.

## 2. Materials & Methods

Lattice arrays consisting of silicon nitride particles on a silicon dioxide substrate were modeled in COMSOL Multiphysics v6.0 [36], a finite-element method simulation software, using the Wave Optics Module [37] (see Supplemental Document Sections S1 and S2 for implementation details and Code 1 for COMSOL file (Ref. [38])). Each particle had a fixed height of 270 nm, and the gap among the particles was held at 100 nm (Fig. 1a). The two-dimensional shape, width, and length of the nanoparticles varied across arrays. The tested shapes included squares, oblongs (non-square rectangles), and L-shapes (squares or oblongs with a removed corner) (Fig. 1b). CPL was used to excite nanostructure arrays.

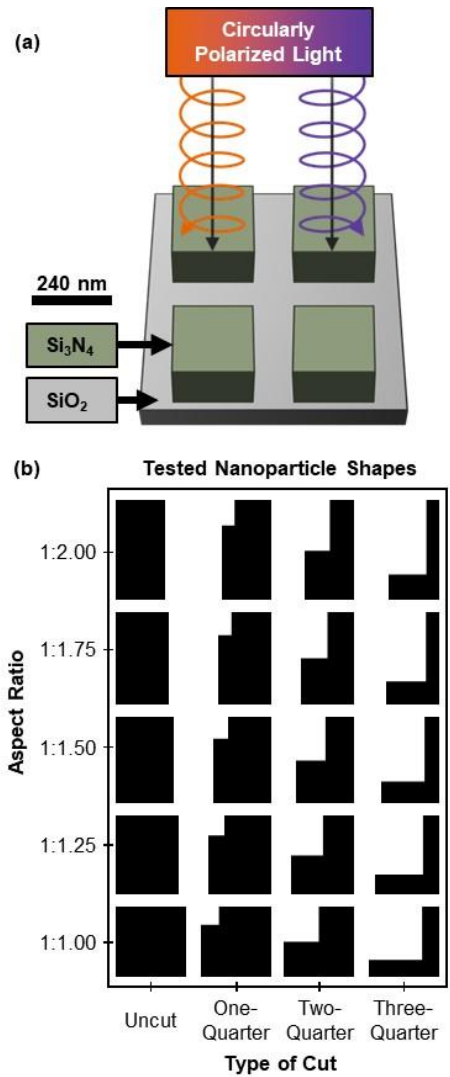

Fig. 1. (a) Schematic of metasurface composed of square-shaped, silicon nitride nanoparticles, arranged periodically on a silicon dioxide substrate with periodic boundary conditions (only four unit cells shown in schematic). (b) Diagram of all the nanoparticles that were tested for this experiment. The type of cut refers to the ratio of the dimensions of the corner cut to the respective dimensions of the overall shape. For a certain cut, the volumes of the nanoparticles were kept the same as aspect ratio changed. Created in BioRender. Poulikakos, L. (2024) BioRender.com/m06q553.

COMSOL Multiphysics v6.0 [36] functionalities were used to obtain near-field data: (i) electric field enhancement values and (ii) net electric flux through each nanoparticle (provided in Dataset 1 (Ref. [39])). Far-field reflectance spectra were also acquired, and also provided in Dataset 1 (Ref. [39]). These spectra were used to predict the structural color output on the CIE 1931 2-degree Standard Observer color space [33,34] (see Supplemental Document Section S3 for color predicting calculations). For a pair of colors, the color-difference was scored using the CIEDE2000 method [40] (see Supplemental Document Section S4 for color-difference calculations). This setup builds on previous work by our team and others [35,41].

The nanoparticles in this work can be categorized either by their aspect ratio or by the modifications made to them (Fig. 1b). All structures had fixed aspect ratios of 1:1.00, 1:1.25, 1:1.50, 1:1.75, and 1:2.00. A modification refers to the portion of a corner removed from a shape, where the size of the cut is determined by a ratio relative to the overall dimensions. For example, a one-quarter cut means the length or width of the cut corner is one-quarter of the total length or width of the shape. This same logic applies to two-quarter and three-quarter cuts, where the size of the removed corner is proportional to the shape's overall dimensions. Exact parameters of all the tested nanoparticles are provided in the Supplemental Document Section S5.1. We note that the results for structures with three-quarter cuts will mostly be provided in the supporting information (see Supplemental Document Section S5.2 for information relevant to three-quarter cut structures). For each modification, the volume of the nanoparticles across aspect ratios was kept constant to control for any potential effects on the arising resonances [42].

## 3. Results & Discussion

We begin our analysis by examining the 1:1.00 aspect ratio nanostructures illuminated with clockwise and counterclockwise CPL (Fig. 2). This uncut, square-shaped structure produces a single resonance peak in reflectance for each CPL orientation (Fig. 2a), in accordance with previous demonstrations [35,41]. This phenomenon is attributed to the orthogonal symmetry of square geometries [43]. Introducing a one-quarter cut into the square structure dampens the resonance peak and results in its splitting into two distinct peaks (Fig. 2b). Further expanding the cut to a two-quarter cut collapses the two resonance peaks into a single peak; although, the amplitude remains reduced and does not return to the level observed in the no cut structure (Fig. 2c).

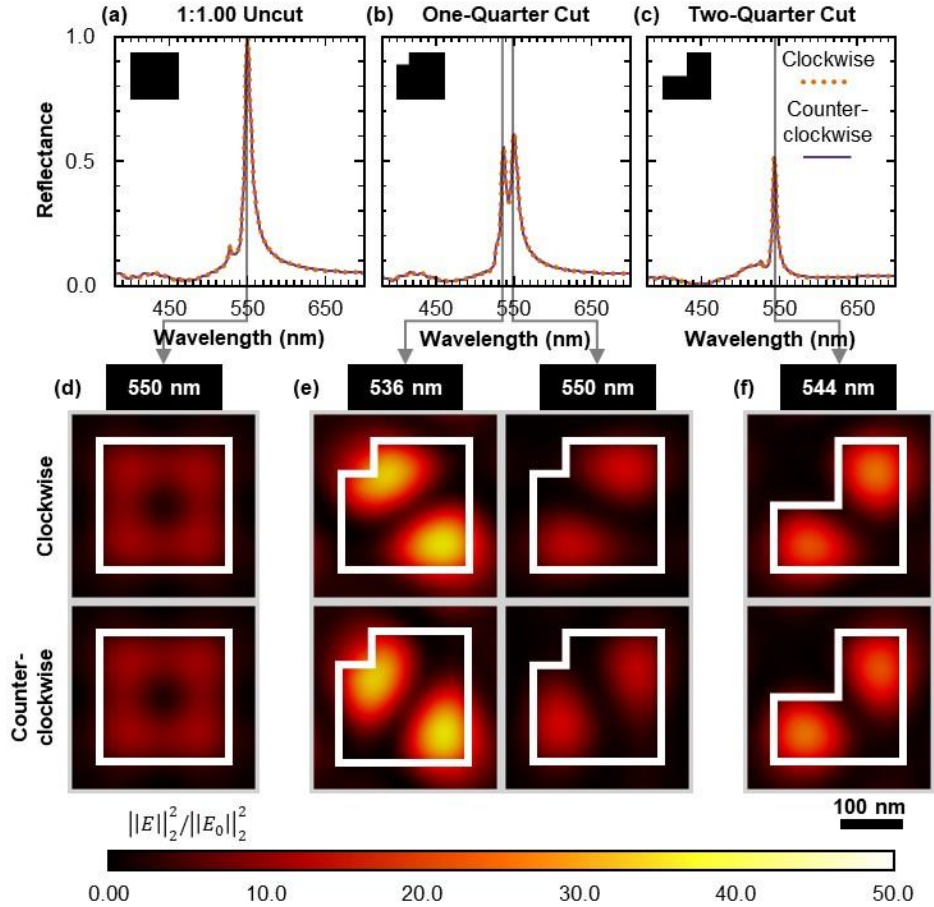

Fig. 2. (a-c) Reflectance plots of the 1:1.00 aspect ratio nanoparticle and corresponding one-quarter cut and two-quarter cut L-shaped nanostructure arrays. All three graphs plot spectra resultant from exciting the nanostructure arrays with broadband – 380 nm through 700 nm – clockwise (orange, dotted) or counterclockwise (purple, solid) CPL. The responses are identical between both polarizations, i.e., the plots overlap. Solid line arrows descending through the peaks of the spectra point to the respective (d-f) near-field electric field enhancement plots of the nanoparticles at the resonance wavelengths. The white outlines delineate the boundaries of the nanostructures under clockwise (top row) or counterclockwise (bottom row) illumination, respectively. The color bar represents the value of electric field enhancement. $E$ is the local electric field vector, $E_0$ is the incident electric field vector, and $||\cdot||_2$ is the Euclidean norm of a vector.

Near-field examinations of the electric field enhancement reveal that the 1:1.00 uncut structure supports an electric quadrupole for both orientations of CPL (Fig. 2d). When a one-quarter cut is introduced, each quadrupole splits into two electric dipoles (Fig. 2e). This behavior is consistent with the fundamental nature of CPL, which is a superposition of two orthogonal linearly polarized light vectors that are quarter-wave phase shifted from one another [1]. Each linear vector contributes to a dipole; when superimposed, these dipoles form a quadrupole. The introduction of a one-quarter cut to the square-shaped structure violates the symmetry, which forces each linear vector to resonate at a distinct wavelength: 536 nm and 550 nm. The spatial distribution of the 536 nm dipoles places one of the resonance nodes at the cut corner. When the cut is expanded to the two-quarter cut, the structure suppresses the electric dipole resonances at 536 nm (Fig. 2f).

These near-field manipulations are corroborated in the far field. The electric quadrupoles in the 1:1.00 uncut structure exhibited the highest amplitude at the resonance wavelength in the reflectance spectra among all three structures. Upon splitting each quadrupole into two dipoles with the one-quarter cut, the far-field spectra evidenced this change with the appearance of two distinct peaks – each with an amplitude nearly halved. When one of the dipoles was suppressed by the two-quarter cut, one peak vanished from the far-field spectra. The remaining peak retained an amplitude approximately half that of the original quadrupole resonances, though with a noticeable blue shift in the resonance wavelength. This blue shift can be attributed to the change in periodicity resulting from the corner cuts [35,41,44]. Similar phenomena were reported in our previous study investigating the destabilization of high-order resonances under linearly polarized light illumination [41]. This phenomena can also be observed in the results of Ref. [45].

The orthogonal symmetry inherent to the 1:1.00 square structure results in identical far-field responses for both clockwise and counterclockwise CPL, even when modifications are made to transform the squares into chiral L-shaped structures. To achieve differential far-field responses, we shift our focus to oblong structures. As previously mentioned, CPL is a superposition of two orthogonal linearly polarized light vectors [1,41]. Oblongs, by definition, lack symmetry along orthogonal axes, leading to interactions with CPL that generate two distinct peaks in the reflectance spectrum (Fig. 3a). However, since the oblong itself is not a chiral shape, these two reflectance peaks overlap perfectly [23].

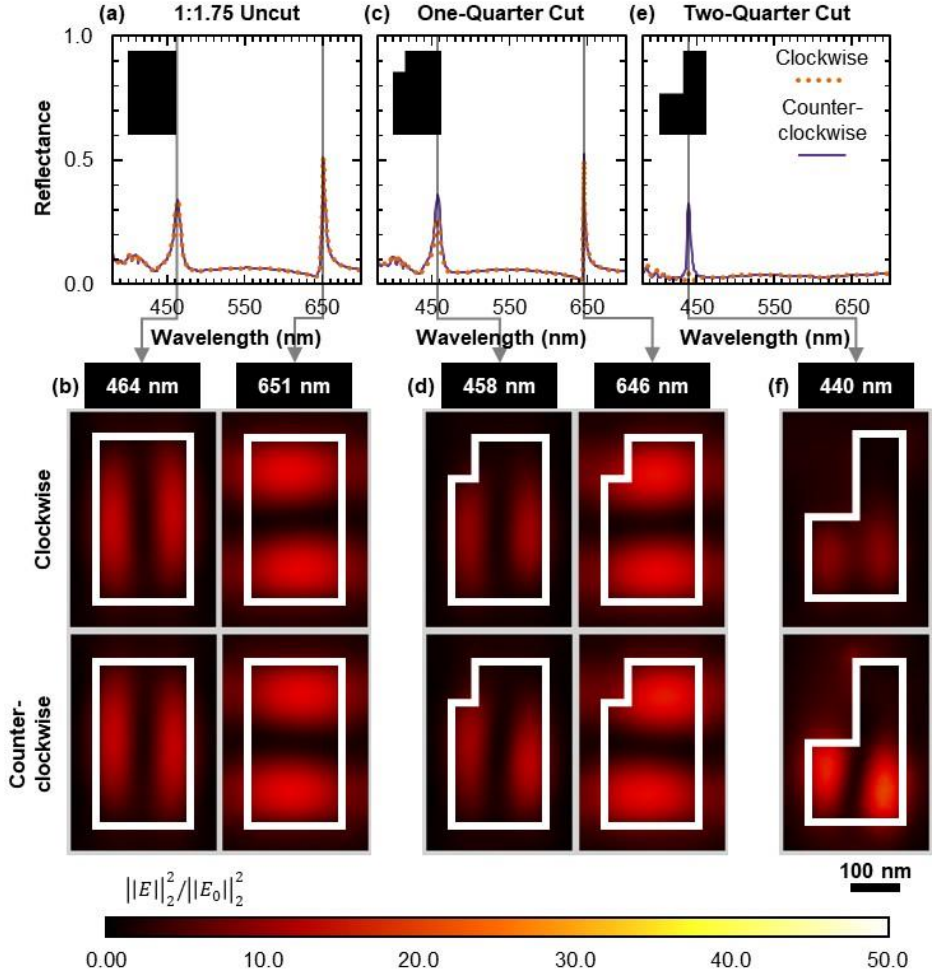

Fig. 3. (a,c,e) reflectance plots of rectangle-shaped and corresponding one-quarter cut and two-quarter cut L-shaped nanostructure arrays. All three graphs plot spectra resultant from exciting the nanostructure arrays with broadband – 380 nm through 700 nm – clockwise (orange, dotted) or counterclockwise (purple, solid) CPL. Solid line arrows descending through the peaks of the spectra point to the respective (b,d,f) near-field electric field enhancement plots of the nanoparticles at the resonance wavelengths. The white outlines delineate the boundaries of the nanostructures under clockwise (top row) or counterclockwise (bottom row) illumination, respectively. The color bar represents the value of electric field enhancement. $E$ is the local electric field vector, $E_0$ is the incident electric field vector, and $|| \bullet ||_2$ is the Euclidean norm of a vector.

An examination of the electric near-field enhancement of the 1:1.75 oblong structure revealed the presence of an electric dipole at each resonance peak for both CPL orientations (Fig. 3b). This observation parallels the behavior observed in the one-quarter cut 1:1.00 square structure, where two electric dipole resonances were detected for each CPL orientation. This led us to investigate whether similar cuts in a rectangular structure could extinguish one of the electric dipole resonances.

Upon introducing a one-quarter cut to the 1:1.75 oblong structure, two resonance peaks per polarization were still observed in the reflectance spectra (Fig. 3c). However, the spectra no longer exhibited a perfect overlap between the two CPL orientations. This discrepancy can be attributed to the two-dimensional-chiral nature of the resulting L-shaped structure [23,46]. Specifically, one pair of resonance peaks was centered at 458 nm, while the other pair was centered at 646 nm. At the 458 nm peaks: the amplitude difference between clockwise and counterclockwise polarizations was 0.09, with the counterclockwise polarization exhibiting the higher amplitude. At the 646 nm peaks: the amplitude difference was 0.03, with the counterclockwise polarization exhibiting the higher amplitude. Thus, the one-quarter cut induced a more pronounced far-field dampening effect on the clockwise polarization. Near-field analysis of the resonance wavelengths for both polarizations confirmed the presence of electric dipoles (Fig. 3d). Notably, the 646 nm resonances displayed nodes positioned much closer to the cut corner compared to the 458 nm resonances. This proximity suggests that the 646 nm resonances are more susceptible to dampening if the cut is expanded to encroach upon the resonance nodes.

In agreement with this hypothesis, the two-quarter cut rectangle exhibited only one resonance peak per polarization in the reflectance spectra (Fig. 3e). This finding corresponds to the detection of a single electric dipole per polarization in the near-field plots (Fig. 3f). The spatial distribution of the nodes corresponds with the shorter-wavelength resonances in the uncut and one-quarter cut 1:1.75 oblong structures, confirming the suppression of the longer-wavelength resonance. The remaining resonance showed a 0.27 amplitude difference between clockwise and counterclockwise resonance peaks in the reflectance spectra, with the counterclockwise polarization showing the higher amplitude. Therefore, the differential far-field dampening effect persisted and became more pronounced.

We aimed to identify the specific aspects of our system contributing to the observed differential response. We start with a qualitative analysis of the electric field vectors through the 1:1.75 uncut, one-quarter cut, and two-quarter cut structures at the shorter-wavelength resonance (Fig. 4a-c). The uncut structure showed a mirrored behavior in the electric field vectors between the clockwise and counterclockwise CPL excitations (Fig. 4a). The distribution of the electric field vectors in the one-quarter cut structure is slightly symmetry-broken (Fig. 4b). Under clockwise polarization, electric field vectors appeared to converge to the cut corner. Meanwhile, under counterclockwise polarization, electric field vectors converged to the adjacent, uncut corner. This elicited a non-mirrored response, which could begin to explain the differential result of the reflectance spectra for this structure. Similarly, the two-quarter cut structure elicited a non-mirrored response (Fig. 4c). This response was more prominent than in the one-quarter cut structure. Notably, there seemed to be a greater flux of the electric field vectors under clockwise polarization along the in-plane x-axis, which is parallel to the longer, thinner arm of the L-shape. Again, this could provide an explanation why this structure had a differential response in its reflectance spectra. Moreover, what appears to be a greater in-plane flux of the electric field for the two-quarter cut structure than in the one-quarter cut structure could explain the differences in differential amplitudes of their respective reflectance spectra.

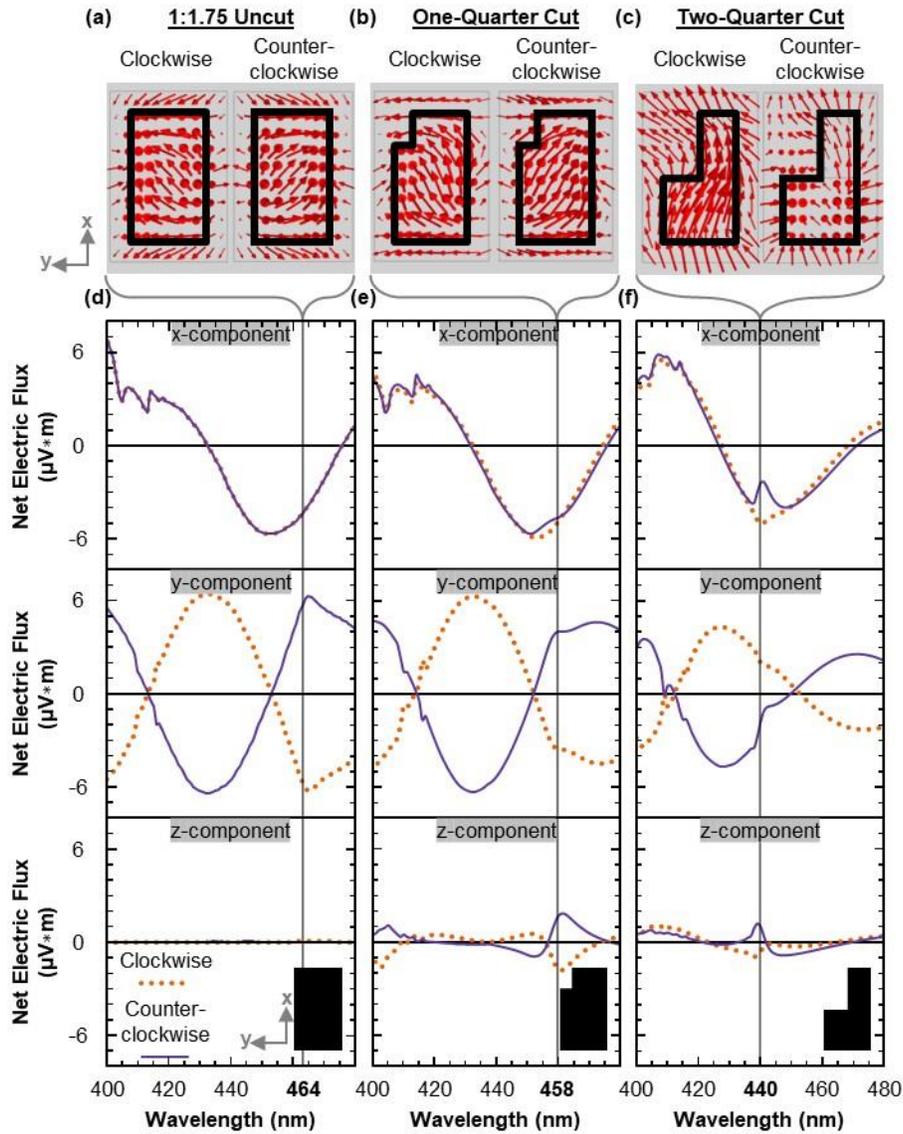

Fig. 4. (a-c) Electric field vectors (red arrows) at the resonant wavelength of the (a) 1:1.75 uncut and corresponding (b) one-quarter cut and (c) two-quarter cut structures. The black outline denotes the nanoparticles' boundaries. (d-f) Net electric flux graphs of the structures from (a-c). (Top) x-component of the net electric flux; (middle) y-component of the net electric flux; (bottom) z-component of the net electric flux. The x- and y-components are aligned with the in-plane scattering. The z-component is aligned with the out-of-plane scattering. All graphs plot net electric fluxes resultant from exciting the nanostructure arrays with clockwise (orange dotted line) or counterclockwise (purple, solid) CPL. Solid, faint gray lines descending through the peaks of the plots point to the respective resonance wavelengths on the horizontal axes. Positive values for the net electric flux represent greater outflux of the electric field lines from the nanostructure, vice versa for the negative values of the net electric flux.

Motivated by this observation, we analyzed the net electric flux for the x-, y-, and z-components of the 1:1.75 oblong structure and its corresponding corner cuts (Fig. 4d-f). The x- and y-components are in-plane with the nanostructure, while the z-component is out-of-plane.

In the uncut rectangular structure, the net electric flux for the x-components showed a perfect overlap between clockwise and counterclockwise CPL (Fig. 4d, top). This result is consistent with the achiral nature of an oblong structure [23]. The y-components were mirror images of each other (Fig 4d, middle). This behavior aligns with the definition of CPL as a superposition of two orthogonal linearly polarized light vectors, where the distinction between clockwise and counterclockwise polarizations is a relative 180-degree phase shift between a linear vector in clockwise CPL compared to the same vector in counterclockwise CPL [1,41]. In our experiments, this vector corresponds to the one aligned with the defined y-axis. The net electric flux for the z-components of both polarizations was found to be two orders of magnitude smaller than the x- and y-component results, appearing nearly zero (Fig 4d, bottom). This indicates that the electric field lines entering and leaving the nanostructure are approximately equal. A positive net flux value would have indicated a greater outflux of electric field lines from the nanostructure, while a negative flux value would have indicated a greater influx of electric field lines into the nanostructure [47,48].

Introducing the one-quarter cut to the structure disrupts the net electric flux symmetry for both CPL orientations (Fig. 4e). In the x-components (Fig. 4e, top), fluxes decouple at the 458 nm resonances. The y-components (Fig. 4e, middle) lose their mirror symmetry, showing unequal flux magnitudes. In the z-components (Fig. 4e, bottom), clear, oppositely handed peaks emerge at the 458 nm resonances. The positive flux in the counterclockwise polarization at the resonance wavelength may explain its higher amplitude in the reflectance spectrum. These z-component flux differences likely stem from the decoupling and loss of mirror symmetry in the x- and y-components, which may be redirecting the electric field lines to scatter unequally out-of-plane between both orientations of CPL. We note that though this analysis only focuses on the shorter wavelength resonance, a similar decoupling of the fluxes was also seen at the longer wavelength resonance (see Supplemental Document, Section S5.3 for full wavelength domain results).

Expanding to the two-quarter cut in the rectangular structure continues to disrupt the net electric flux symmetry across all components (Fig. 4f). This configuration suppresses the longer-wavelength resonances, leaving only a 440 nm resonance. In the x-components (Fig. 4f, top), there is an obvious decoupling at 440 nm where the clockwise peak points downwards and the counterclockwise peak upwards, showing the largest magnitude difference so far. The magnitude of the clockwise peak is greater than the counterclockwise peak. As a result, we expect to see more electric field lines along the x-axis under clockwise polarization, which matches the distribution of near-field electric vectors in Fig. 4c. The y-components (Fig. 4f, middle) displayed a larger 'kink' in the clockwise polarization. In the z-components (Fig. 4f, bottom), symmetry violations persist, with the clockwise flux tending to zero from the negative and the counterclockwise flux showing a prominent positive peak. Combined with the increased symmetry disruptions in the x- and y-components, these results further support the differential redirection of electric field lines to the z-direction. Altogether, this could explain the large differential response in the reflectance spectra at 440 nm, where the counterclockwise polarization shows a +0.27-amplitude difference.

The observed differential net flux response arises from an unequal distribution of electric field vectors entering or leaving the nanoparticle along a specific axis [47,48]. In our all-dielectric system, we suspect this effect emerges at the resonance wavelength due to electron displacement driven by a polarization effect from the incoming electromagnetic wave [48,49]. This displacement generates an internal electric flux source [49], forming an electric dipole [50], as shown in Fig. 2 and Fig. 3. Although other modal resonances can occur when the nanoparticle sizes approach the wavelength [41,51], this was not the case for our tested nanoparticles in the visible light regime.

When mirror symmetry of the nanoparticle is broken, we hypothesize the electron displacement will vary upon excitation with clockwise and counterclockwise CPL, leading to an asymmetric internal flux between both polarizations. This internal flux asymmetry interacts with the external field, explaining the differential net electric flux responses in Fig. 4, which only appeared when corner cuts were introduced to the nanoparticle. Similarly, the 1:1.00 structures also showed differential net electric flux responses when their mirror symmetry was disrupted by corner cuts (see Supplemental Document Section S5.3 for net electric flux results for all structures).

However, while the 1:1.00 structures did not produce a differential far-field response, the 1:1.75 structures did. Both structures lose mirror symmetry with the corner cuts, but only the 1:1.75 configuration becomes two-dimensionally chiral. This suggests that a differential far-field response only occurs when two conditions are met: (i) the nanoparticle shape is chiral and (ii) the near-field net electric flux is asymmetric. Therefore, although near-field net electric flux asymmetry may not directly cause far-field effects, it likely serves as an indicator of far-field behavior, with similar underlying factors influencing both near-field and far-field responses.

Differential far-field responses between clockwise and counterclockwise CPL can be translated into distinguishable colors. We calculated the predicted colors from the tested structures in their uncut and cut configurations, and plotted them on the CIE 1931 2-degree Standard Observer color space [33,34]. Due to the orthogonal symmetry of the 1:1.00 square and corresponding L-shaped structures, their far-field responses and predicted colors were identical across CPL orientations (see Supplemental Document Section S5.4 for colorimetry results for all structures). In contrast, the 1:1.75 oblong and corresponding L-shaped structures, lacking symmetry along orthogonal axes, produced distinct color signals once mirror symmetry was broken to achieve chirality (Fig. 5a-c). The two-quarter cut structure exhibited the largest color difference (Fig. 5c), which was further confirmed by the CIEDE2000 color-difference scores [40] (Fig. 5d). This significant differentiation persisted across other oblong structures with varying aspect ratios (Fig. 5e) (see Supplemental Document Section S5.4 for colorimetry results for all structures). The two-quarter cut configuration simplifies CPL differentiation to a single peak location in the far-field spectra, further increasing color differences between the two polarizations, which was further supported by the descriptive statistics (see Supplemental Document Section S5.5 for colorimetry descriptive statistics).

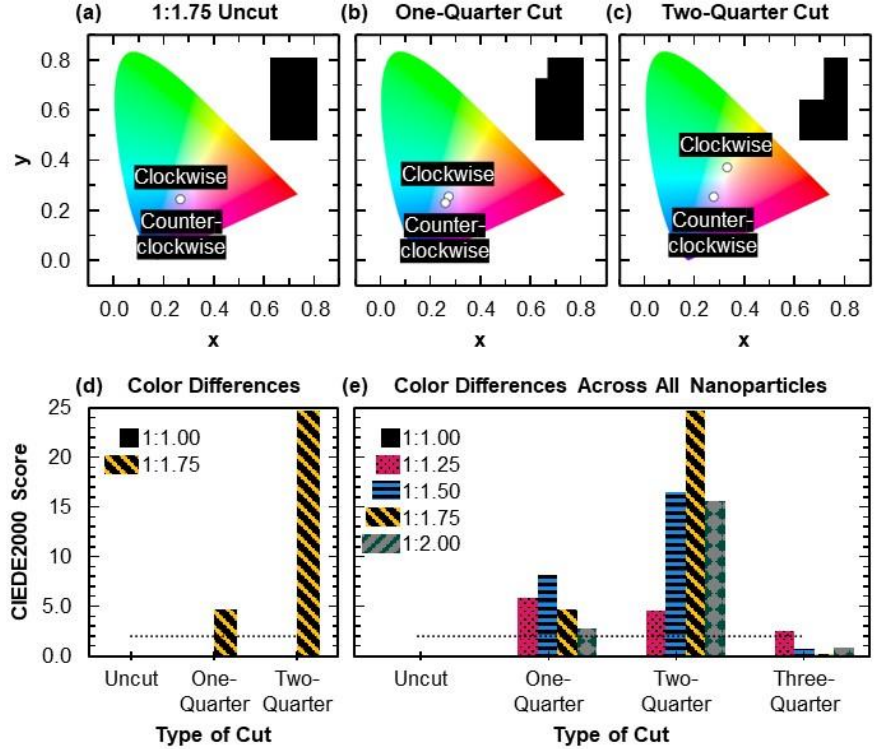

Fig. 5. (a-c) CIE 1931 2-degree Standard Observer color space plots for the (a) uncut, (b) one-quarter cut, and (c) two-quarter rectangular structures. (d) The color difference scores, calculated using the CIEDE2000 method, between the clockwise and counterclockwise polarizations for the square and 1:1.75 rectangle. The square presented no color difference, so the score is zero for all cut types. (e) The CIEDE2000 color difference score for all the nanoparticles shown in Fig. 1b. The horizontal dotted line in (d) and (e) indicates the minimum CIEDE2000 score (= 2.3) needed to establish that two colors are perceptibly different.

## 4. Conclusion

From the numerical studies presented in this work, we hypothesize that the net electric flux through an element of an all-dielectric metasurface may correlate with the differentiation between clockwise and counterclockwise CPL in the far-field spectrum. By designing metasurface elements such that the electric field lines of one polarization disperse along the in-plane axes and those of the other polarization scatter along the out-of-plane axis, it may be possible to maximize this differentiation within the studied design space. This hypothesis, which developed from inducing a one-quarter or two-quarter cut in the nanostructure elements of our metasurfaces, leads to differences in net electric flux between clockwise and counterclockwise CPL illumination. These differences are apparent in chiral L-shaped structures but absent in the achiral square or oblong structures, suggesting a possible correlation with far-field responses. Thus, we propose that optimizing the nanostructures for net electric flux could enhance the differentiation of CPL orientations; if limited to the visible regime, then potentially providing a color sensor with maximal color difference scores between clockwise and counterclockwise CPL.

Our findings are currently correlative, and further study is required to assess the strength of this correlation. The dataset is limited to three rectangular L-shaped structures that suppress resonances (see Supplemental Document Section S5 for results), necessitating a broader exploration to confirm these preliminary results. The modelled structures did not have smoothed corners to mimic nanofabrication artifacts, which may decrease the near-field enhancement values due to a reduction in corner sharpness [52]. Previous studies showed that fabrication artifacts result in a reduced amplitude and expanded linewidth of the reflectance peaks, but generally align with the numerical models [35,53]. As a result, there may be effects on the perceived colors between the two polarizations, as our colorimetry results rely on differential peak amplitudes at the same resonance wavelength. Nevertheless, if the correlations between the net electric flux and far-field measurements hold, optimized structures can be developed to achieve a more pronounced differential amplitude between the two polarizations. This would compensate for the expected decrease in amplitude upon fabrication.

To further investigate this correlation, several avenues can be pursued. First, exploring a larger parameter space could yield a more robust dataset for statistical analysis. Second, our study kept the cut width and length proportional to the overall structure, but varying these proportions could identify optimal conditions for suppressing one of the two electric dipoles in rectangular structures. Third, while L-shaped structures were used in this study due to their simplicity, they may not be the ideal shape for differentiating between CPL orientations. Investigating how perturbations to the L-shapes affect the net electric flux and far-field measurements could help refine the design. Finally, establishing a causative relationship, rather than a mere correlation, may require machine learning approaches. Inverse design methods using theoretical models – such as finite element methods or rigorous coupled wave analysis – could work but will suffer from increased computational intensity arising from a lack of differentiable solutions [54]. As an alternative, a physics-driven deep learning model could offer a feasible alternative by aligning solutions with Maxwell's equations while continuously measuring near-field effects and far-field responses. This alternative will require the acquirement of a large enough and appropriate dataset to for training a neural network [54].

In summary, this study presents a methodology to simplify the differentiation of CPL orientations from a two-peak to a single-peak problem and hypothesizes a correlation between net electric flux and far-field observations. Future research should aim to validate and strengthen this correlation and explore the potential for a causative relationship using advanced computational techniques.

## 5. Back Matter

### 5.1 Funding

Z.H. and L.V.P. gratefully acknowledge funding from the Arnold and Mabel Beckman Foundation (Beckman Young Investigator Award, Project Number: 30155266). A.M.N. gratefully acknowledges funding from the UC LEADS program.

### 5.2 Disclosures

The authors declare no conflicts of interest.

### 5.3 Data Availability Statement

Data underlying the results presented in this paper are available in Dataset 1, Ref. [39].

### 5.4 Supplementary Document

See Supplement 1 for supporting content.

# Supplemental Document: Chiral colorimetry in lossless all-dielectric metasurfaces linked to net electric flux

Zaid Haddadin[1], Anna My Nguyen[2], and Lisa V. Poulikakos[2,3,*]

[1]Department of Electrical & Computer Engineering, UC San Diego, La Jolla, CA 92093-0021, USA
[2]Department of Mechanical & Aerospace Engineering, UC San Diego, La Jolla, CA 92093-0021, USA
[3]Program of Materials Science & Engineering, UC San Diego, La Jolla, CA 92093-0021, USA
*lpoulikakos@ucsd.edu

**Abstract:** All-dielectric metasurfaces can produce structural colors, but their design criteria are under investigation. This work numerically studies how the two-dimensional shape of nanoparticles affects the colorimetric response under circularly polarized light (CPL) to develop a sensor distinguishing CPL orientations. Using lossless dielectric materials, we achieve far field dichroism by modifying oblong nanoparticles into L-shaped structures through corner cuts. This design suppresses one electric dipole under CPL illumination, leading to differential colorimetric responses. We link these responses to a decoupling effect in the near field net electric flux. Our findings provide design guidelines for all-dielectric, lossless colorimetric sensors of chiral light.

## S1. In-lab Workstation Specifications

Numerical simulations were solved on a Precision 3660 Tower from Dell Technologies [1] using a 64-bit Windows 11 Pro operating system [2]. Device and Windows specifications can be found in Table S1 and Table S2, respectively. The full spec sheet for the Precision 3660 Tower can be found in Ref. [3].

**Table S1. In-lab Device Specifications for Precision 3660 Tower from Dell Technologies**

| Processor | 12th Gen Intel(R) Core(TM) i7-12700 2.10 GHz |
|---|---|
| Installed RAM | 32.0 GB (31.7 GB usable) |
| System type | 64-bit operating system, x64-based processor |

**Table S2. In-lab Windows 11 Specifications**

| Edition | Windows 11 Pro |
|---|---|
| Version | 23H2 |
| Installed on | 10/10/2023 |
| OS Build | 22631.4169 |

## S2. COMSOL Multiphysics v6.0 Set Up

The COMSOL Multiphysics v6.0 base software [4] and the Wave Optics Module [5] were used for this work. The COMSOL file was created with the "Electromagnetic Waves, Frequency Domain" layer in the three-dimensional (3D) space. The file was created with an empty study. An unsolved simulation file for the 1:1.75 Oblong Two-Quarter Cut structure (Main Text, Fig. 1b) is provided as Code 1 in Ref. [6]. However, some parameters in the file are redundant – a consequence of trial & error testing to build up a working file. As such, we highlight the relevant portions of the file in the sections below, which follow the naming conventions of the default groupings generated by COMSOL.

*Global Definitions*

Three parameter layers were created: (i) Model Parameters, (ii) Nanostructure Parameters, and (iii) Optics Parameters. The attributes and entries of these parameter layers are provided in Tables S3, S4, and S5, respectively.

| Table S3. Model Parameters | | | |
|---|---|---|---|
| **Name** | **Expression** | **Value** | **Description** |
| **domain_depth** | lattice_depth_period | 2.8174E-7 m | The depth of the entire domain. |
| **domain_height** | incident_height + transmitted_height + nanostructure_thickness | 4.27E-6 m | The height of the entire domain. This is composed of all parts of the model. |
| **domain_width** | lattice_width_period | 4.1804E-7 m | The width of the entire domain. |
| **incident_height** | 2000 [nm] | 2E-6 m | The height of the domain through which light is incident towards the thin film. |
| **transmitted_height** | 2000 [nm] | 2E-6 m | The height of the domain through which light is transmitted after passing through the thin film. |
| **lattice_width_period** | nanostructure_pitch_width + interresonator_gap | 4.1804E-7 m | The interval of distance between successive repetitions of the nanoresonator along the width of the nanostructure. |
| **lattice_depth_period** | nanostructure_pitch_depth+ interresonator_gap | 2.8174E-7 m | The interval of distance between successive repetitions of the nanoresonator along the depth of the nanostructure. |

| Table S4. Nanostructure Parameters | | | |
|---|---|---|---|
| **Name** | **Expression** | **Value** | **Description** |
| **nanostructure_thickness** | 270 [nm] | 2.7E-7 m | The thickness (or height) of the nanostructure. |
| **nanostructure_pitch_width** | nanostructure_pitch_depth * aspect_ratio_factor | 3.1804E-7 m | The width of the nanostructure. |
| **nanostructure_pitch_depth** | sqrt((170 [nm] * 340 [nm]) / aspect_ratio_factor) | 1.8174E-7 m | The depth of the nanostructure. |
| **interresonator_gap** | 100 [nm] | 1E-7 m | The gap between two adjacent nanostructures. |
| **aspect_ratio_factor** | 1.75 | 1.75 | "This factor is x in 1:x, assuming the shorter end of the |

| | | | | structure will always be the ""1""." |
|---|---|---|---|---|
| **nanostructure_cornercut_depth** | nanostructure_pitch_depth / cornercut_ratio_factor | | 9.0869E-8 m | The depth of the cut that'll happen at the nanostructure's corner. |
| **nanostructure_cornercut_width** | nanostructure_pitch_width / cornercut_ratio_factor | | 1.5902E-7 m | The width of the cut that'll happen at the nanostructure's corner. |
| **cornercut_ratio_factor** | 2 | | 2 | |

| Table S5. Optics Parameters | | | |
|---|---|---|---|
| **Name** | **Expression** | **Value** | **Description** |
| **wavelength_min** | 380 [nm] | 3.8E-7 m | The minimum value the wavelength of light can take for this study. |
| **wavelength_max** | 700 [nm] | 7E-7 m | The maximum value the wavelength of light can take for this study. |
| **wavelength_step** | 1 [nm] | 1E-9 m | The steps taken in incrementing/decrementing the wavelength when running the simulation. |
| **polarisation_rotation_angle** | 90 [deg] | 1.5708 rad | The trigger between CW and CCW CPL |

*Component 1*

This grouping contains further sub-groupings: (i) Definitions, (ii) Geometry 1, (iii) Materials, (iv) Electromagnetic Waves, Frequency Domain, and (iv) Mesh. The relevant contents of each sub-grouping are highlighted in their respective sections, which follows this.

Definitions

Two variable layers and one integration layer were created. The first variable layer, "Electric & magnetic mode field amplitudes", was used to switch between incident clockwise and counterclockwise circularly polarized light (Table S6). The second variable layer, "Incident Electric & Magnetic Field Norms", was used to define the electric & magnetic field enhancement values (Table S7) and relies on the integration layer, "Integration 1 (*IntegrateOverInputPort*)", which defines an area at which to integrate over (Fig. S1).

| Table S6. Electric & magnetic field amplitudes | | | |
|---|---|---|---|
| **Name** | **Expression** | **Unit** | **Description** |
| **E0x** | 1 / sqrt(2) | | The x-component of the electric field amplitude. |
| **E0y** | (1 / sqrt(2)) * i * sin(polarisation_rotation_angle) | | The y-component of the electric field amplitude. |

| Table S7. Incident Electric & Magnetic Field Norms | | | |
|---|---|---|---|
| **Name** | **Expression** | **Unit** | **Description** |

| **IncidentElectricFieldNorm** | IntegrateOverInputPort(ewfd_TE.normE) / (domain_depth * domain_width) | | V/m | |
| --- | --- | --- | --- | --- |
| **IncidentMagneticFieldNorm** | IntegrateOverInputPort(ewfd_TE.normH) / (domain_depth * domain_width) | | A/m | |

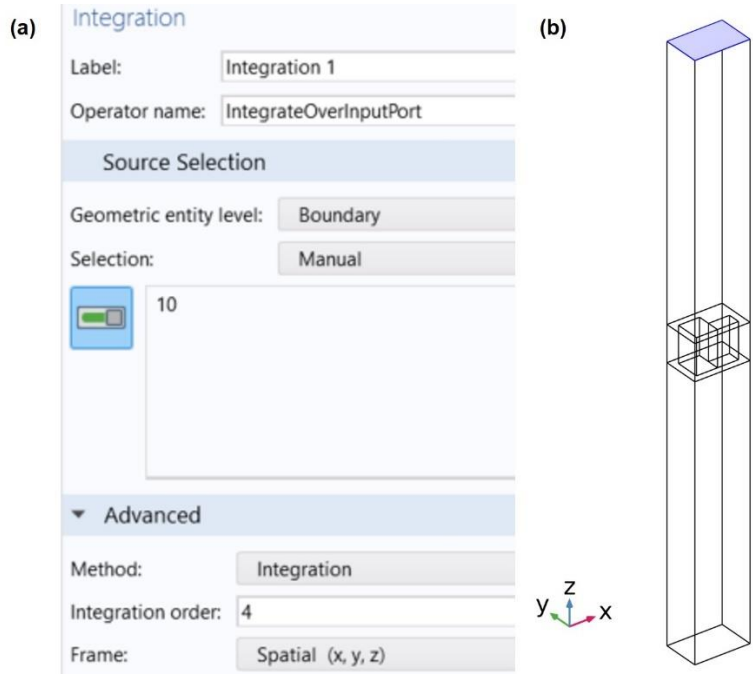

Fig. S1. (a) Parameters used in the "Integration 1" layer. The operator name, "IntegrateOverInputPort", defines the variable name that can be used with other functions to integrate over the selection. In this case, the selection is boundary #10, which refers to the port that defines the incoming electromagnetic wave. (b) A schematic of the entire model with the selected port boundary highlighted in blue.

### Geometry 1

The Geometry 1 grouping has the option to define a length unit, which we set to "nm". Following that three layers were created: (i) a block layer, (ii) a work plane layer, and (iii) an extrude layer. The block layer was used to define the entire working domain of the model (Fig. S2). The work plane layer was used to define the two-dimensional shape of the nanoparticle (Fig. S3). This required creating three sub-layers: two rectangle layers (Fig. S4 and Fig. S5) and one difference layer (Fig. S6). Finally, the extrude layer was used to transform the two-dimensional nanoparticle from the work plane layer to a three-dimensional nanoparticle (Fig. S7).

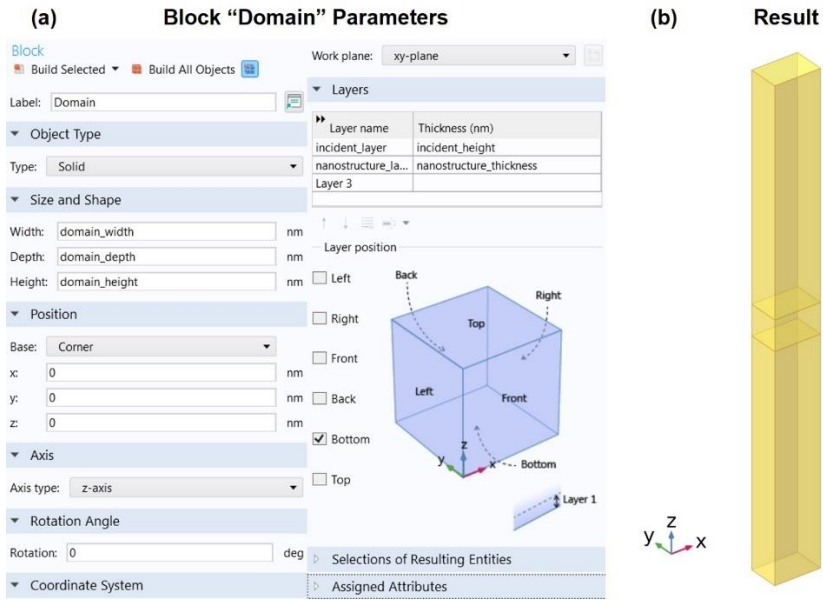

Fig S2. (a) The parameters used in the block layer, named "Domain". The final two sub-sections of this layer, "Selections of Resulting Entities" and "Assigned Attributes", did not contribute to the result as everything remained unchecked within them. (b) The resulting model that is built from the parameter values of the block layer.

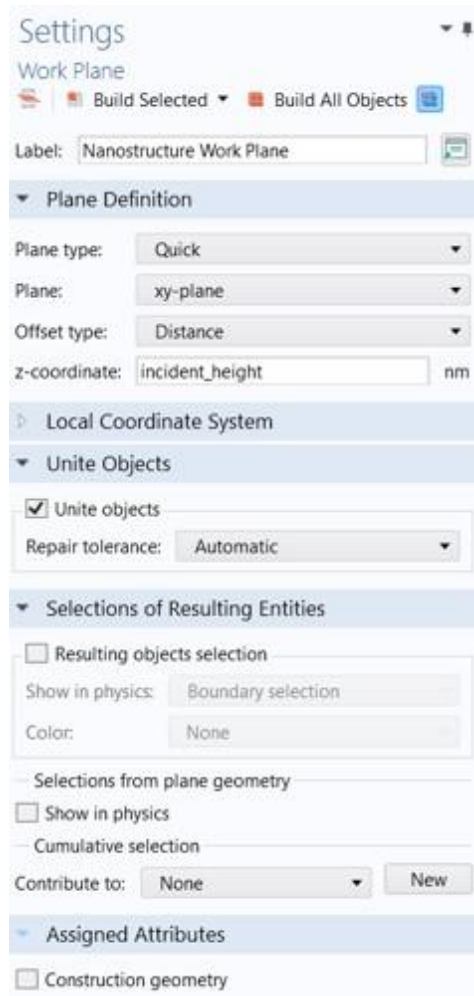

Fig S3. Work plane parameters. This defines the location of the nanostructure in the overall model.

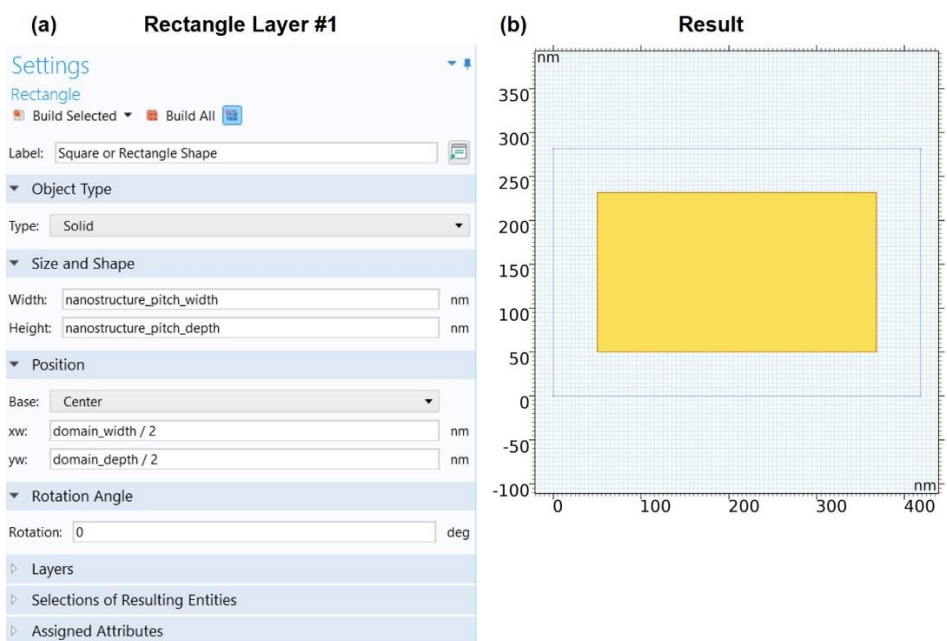

Fig. S4. (a) The first rectangle layer's parameters. This layer was named "Square or Rectangle Shape" as it defines the nanoparticle's two-dimensional shape without corner cuts. (b) The resultant, two-dimensional shape of the nanoparticle.

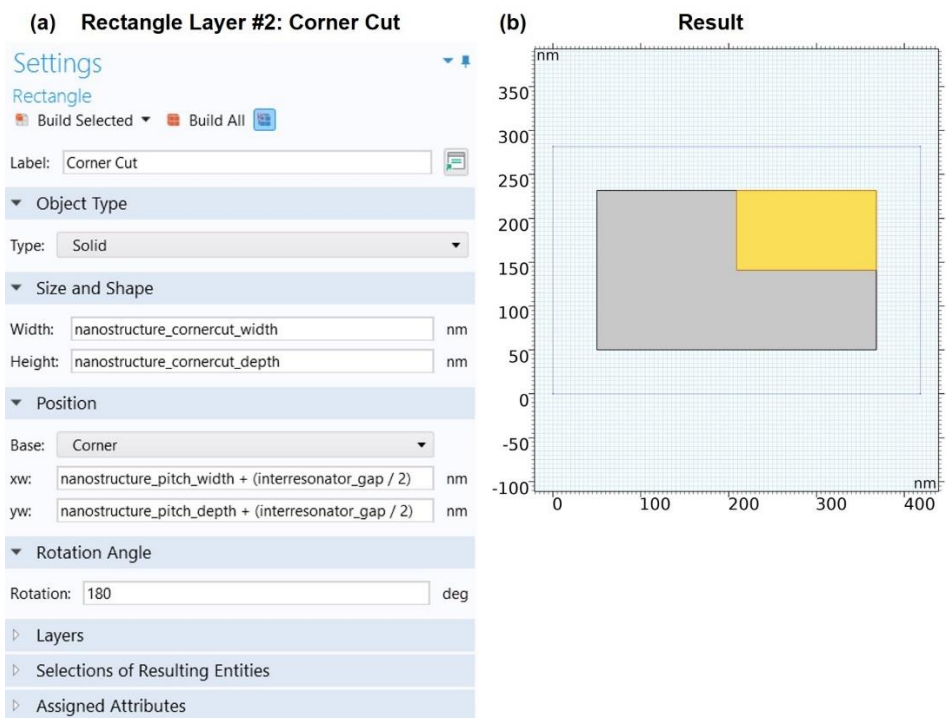

Fig. S4. (a) The second rectangle layer's parameters, which defines the corner cut. (b) The result of the parameters, indicated by the yellow highlight.

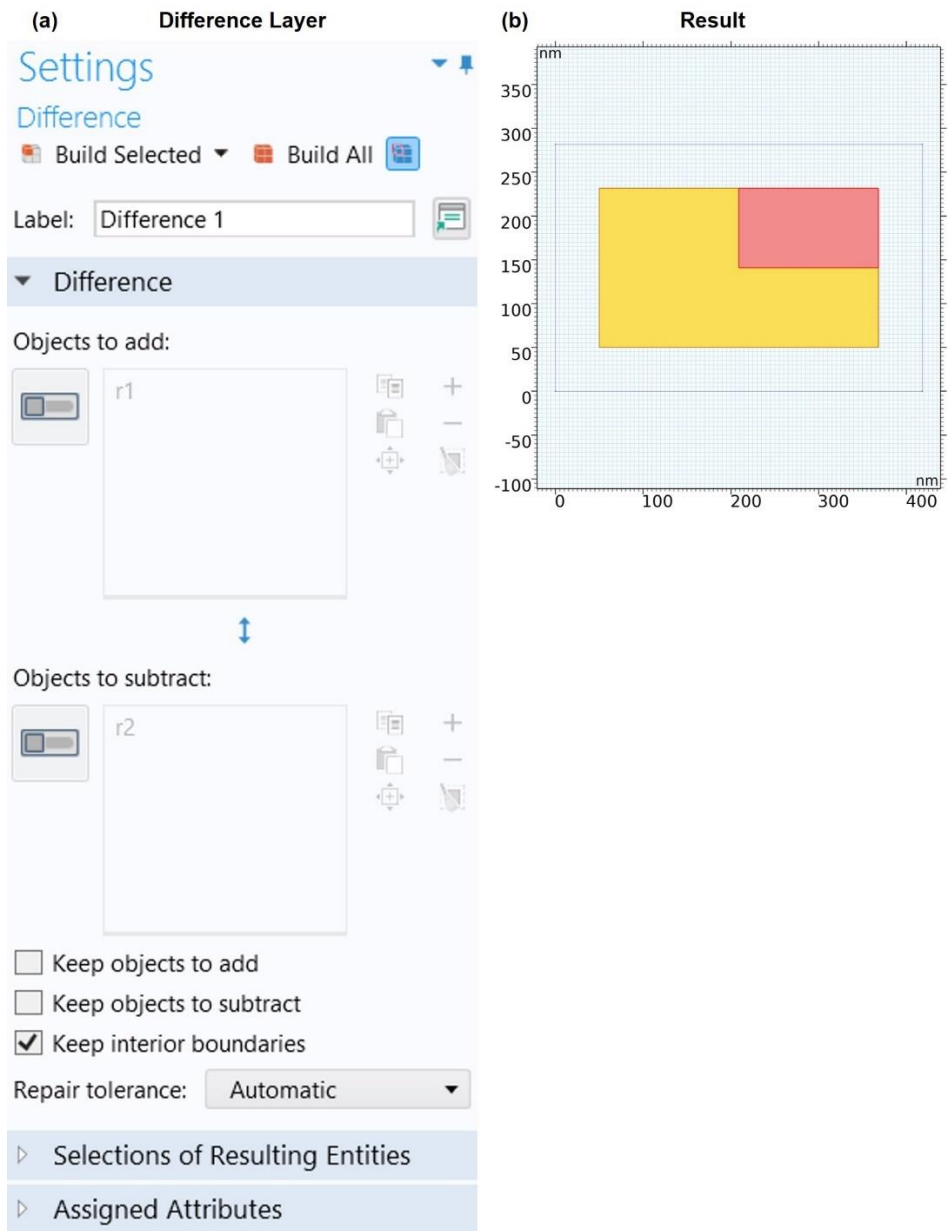

Fig. S6. (a) The parameters of the difference layer. This layer takes the square or oblong object (from the first rectangle layer, labelled "r1" in this figure) and subtracts from it the corner cut object (from the second rectangle layer, labelled "r2" in this figure). (b) The result showing the subtraction of the corner cut object (in red) from the overall object (in yellow).

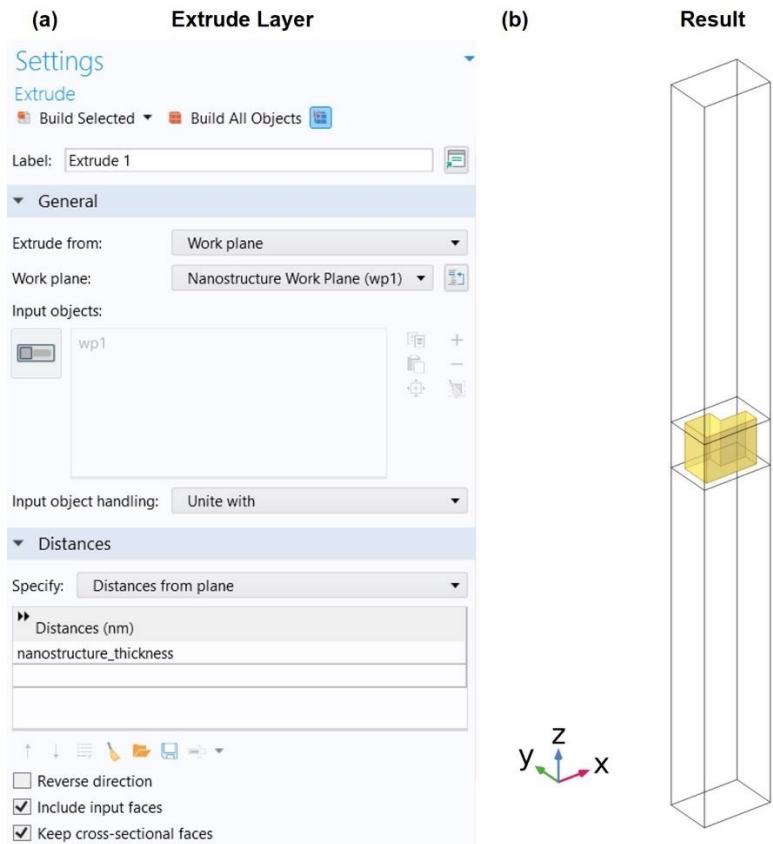

Fig. S7. (a) The parameters used in the extrude layer. "wp1" refers to the work plane layer created in Fig. S3. The screenshot does not show the parameters for five sub-sections: "Scales", "Displacements", "Twist Angles", "Selection of Resulting Entities", and "Assigned Attributes". The unshown sub-sections were not used. (b) The result of the extrude layer, highlighting in yellow the extruded nanoparticle and its location within the model domain.

## Materials

This Materials grouping was where materials were assigned to different portions of the model. Three materials were added using the in-built COMSOL Multiphysics v6.0 Wave Optics Module library: (i) "Air (Ciddor 1996: n 0.23-1.690 μm)" [7], (ii) "Si3N4 (Silicon nitride) (Luke et al. 2015: n 0.310-5.504 μm)" [8], and (iii) "SiO2 (Silicon dioxide, Silica, Quartz) (Ghosh 1999: α-Quartz, n(o) 0.198-2.0531 μm)" [9]. Fig. S8 shows the distribution of these three materials throughout the model.

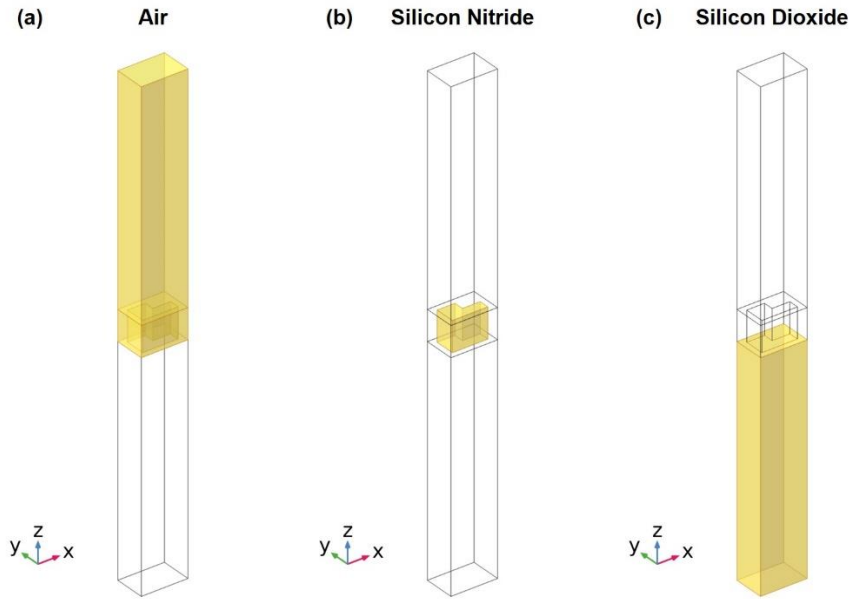

Fig. S8. Distribution of (a) air, (b) silicon nitride, and (c) silicon dioxide across the model domain. The parts of the model that are highlighted in yellow indicate where the material was applied.

### Electromagnetic Waves, Frequency Domain

The Electromagnetic Waves, Frequency Domain grouping has the option to define the formulation of the electromagnetic wave. We set this to "Full field" while having "All domains" selected. Following that, we defined parameters within two of three default sub-layers: "Wave Equation, Electric 1" (Fig. S9) and "Initial Values 1" (Fig. S10). We ignore the "Perfect Electric Conductor 1" layer. Then, we created two port layers (Fig. S11 and Fig. S12) and two periodic condition layers (Fig. S13).

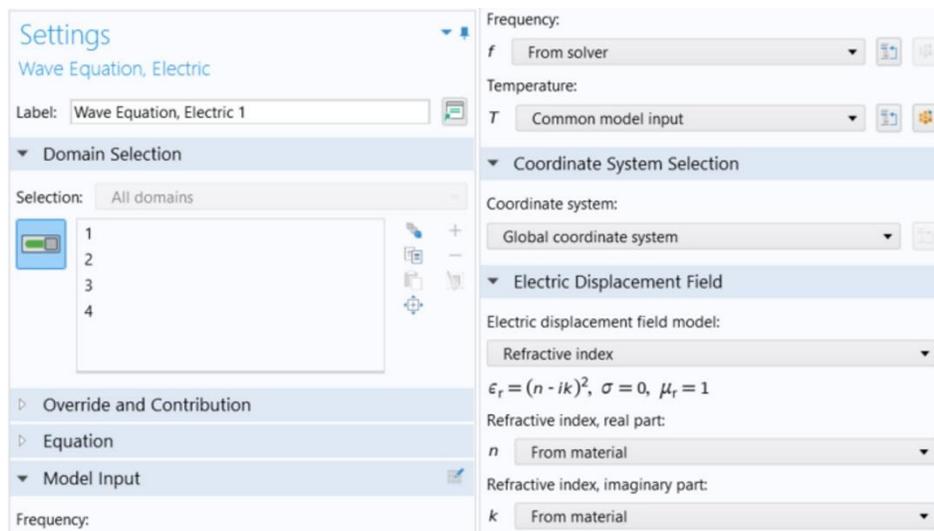

Fig. S9. Parameters used for the Electromagnetic Waves, Frequency Domain grouping.

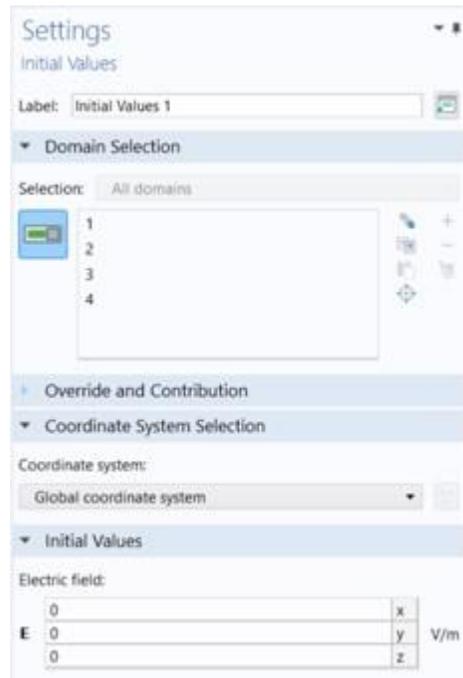

Fig. S10. Parameters for the "Initial Values 1" layer.

Fig. S11. Parameters for the "Port 1" layer.

Fig. S12. Parameters for the "Port 2" layer.

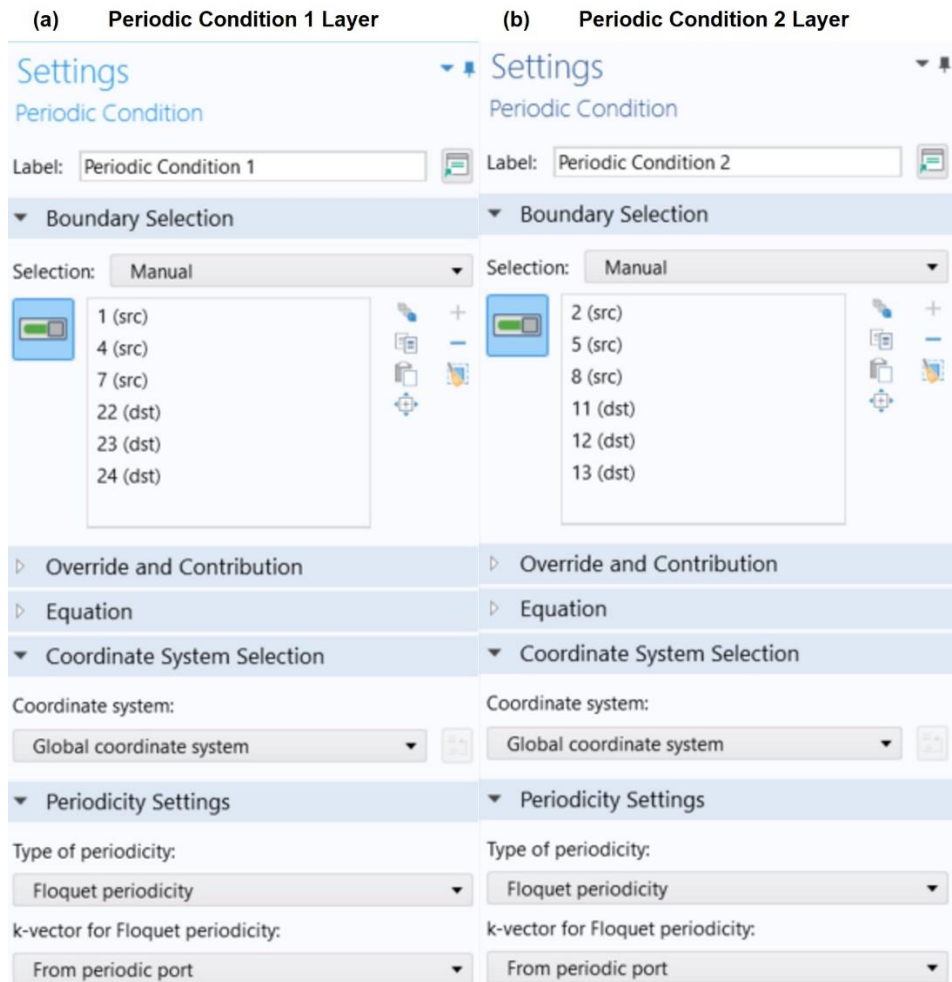

Fig. S13. The parameters used in the (a) first and (b) second periodic condition layers.

Mesh

The model was rendered using a physical-controlled mesh with the default "Normal" element size (Fig. S14).

Page **14** of **31**

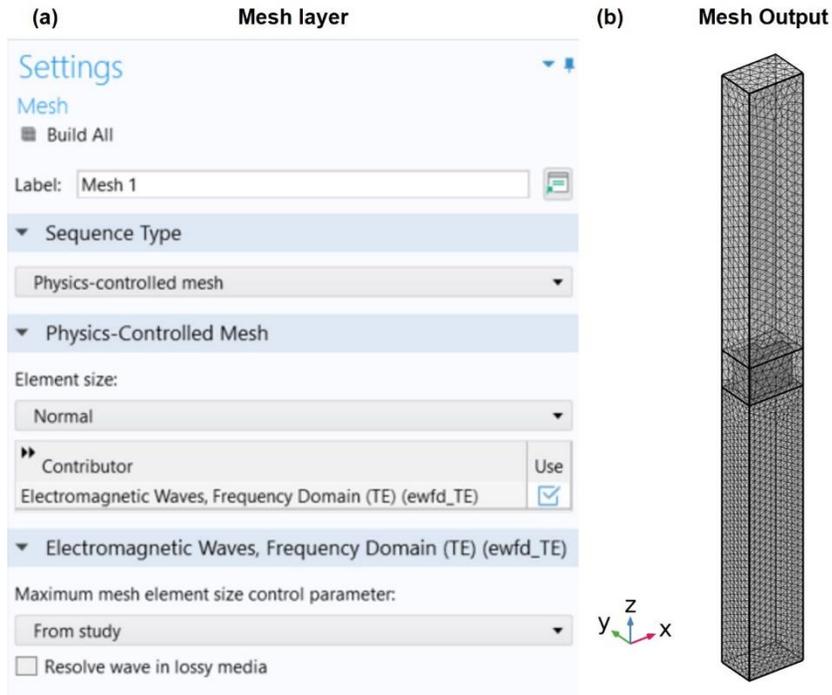

Fig. S14. (a) The parameters used in the mesh layer. (b) The output of rendering the mesh.

## S3. Predicted Structural Color Outputs

The reflectance spectra generated throughout this work were used to predict the structural color output on the CIE 1931 2-degree Standard Observer color space [10,11]. The XYZ tristimulus values quantify the amount of red, green, and blue recorded by the human eye [Eq. (1)-(3)].

$$X = k \sum_\lambda \phi_\lambda(\lambda) \bar{x}(\lambda) \Delta\lambda \qquad (1)$$

$$Y = k \sum_\lambda \phi_\lambda(\lambda) \bar{y}(\lambda) \Delta\lambda \qquad (2)$$

$$Z = k \sum_\lambda \phi_\lambda(\lambda) \bar{z}(\lambda) \Delta\lambda \qquad (3)$$

The spectral distribution for a wavelength $\lambda$ is denoted by $\phi_\lambda$ [12], where $\lambda$ ranged from 380 nm to 700 nm, inclusive. $\bar{x}$, $\bar{y}$, and $\bar{z}$ represent the colour matching functions of a standard colorimetric observer (Fig. S15) [13].

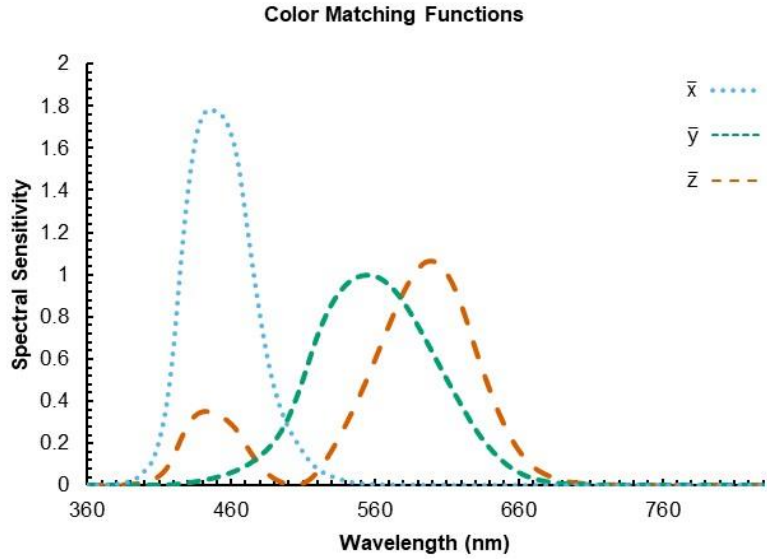

Fig. S15. Color matching functions for the CIE 1976 2-degree Standard Observer color space.

Color matching functions define how an observer's visual system responds to red, green, and blue primaries [10,11]. The functions were derived by interpolating empirical measurements and are used to calculate tristimulus values [10–12]. $\Delta\lambda = 1$ nm denotes the step-size used for the wavelength measurements [12]. The constant $k$ is chosen such that $Y = 100$ for objects with reflectance equal to 1.0 across all measured wavelengths [Eqn. (4)] [12].

$$k = \frac{100}{\sum_\lambda S(\lambda)\bar{y}(\lambda)\Delta\lambda} \quad (4)$$

$S(\lambda)$ is the spectral distribution of a selected illuminant [12]. In this work, the CIE standard illuminant D65 was chosen (Fig. S16) [14,15]. This illuminant represents average daylight in Northern and Western Europe, with a color temperature of 6504 K [14,15]. Because it is an artificial illuminant, translation of our nanostructure arrays into the physical world is expected to yield a different colorimetric response under real light sources [16]. However, due to the challenges in obtaining tristimulus values for reflective media, such as our nanostructure arrays, we had to select a standard illuminant [16]. The choice of D65 is arbitrary.

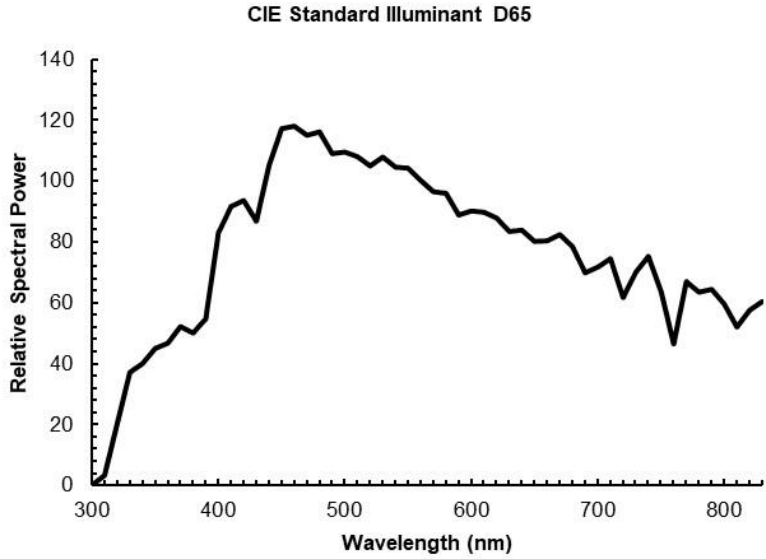

Fig. S15. Relative spectral power distribution of the CIE standard illuminant D65.

By normalizing the XYZ tristimulus values, the chromaticity coordinates $(x, y)$ can be derived [Eqn. (5)-(6)] [17]. The chromaticity coordinates allow for a two-dimensional representation of the color specified by the XYZ tristimulus values in the CIE 1931 2-degree Standard Observer Space [17,18].

$$x = \frac{X}{X+Y+Z} \tag{5}$$

$$y = \frac{Y}{X+Y+Z} \tag{6}$$

**S4. CIEDE2000 Color Difference Evaluation**

The structure of the CIEDE2000 formula is presented in [Eqn. 7] [19]. There are five steps, outlined in the following sub-sections, to calculate the CIEDE2000 color difference between any two given colors represented by the XYZ tristimulus values [20].

$$\text{CIEDE2000} = \left[ \left( \frac{\Delta L'}{k_L S_L} \right)^2 + \left( \frac{\Delta C'}{k_C S_C} \right)^2 + \left( \frac{\Delta H'}{k_H S_H} \right)^2 + R_T \left( \frac{\Delta C'}{k_C S_C} \right) \left( \frac{\Delta H'}{k_H S_H} \right) \right]^{0.5} \tag{7}$$

*Step 1: Converting XYZ to L\*a\*b\**

Let $(X_1, Y_1, Z_1)$ and $(X_2, Y_2, Z_2)$ represent XYZ tristimulus values for a pair of colors. Each of these colors can be converted to the 1976 CIE (L*a*b*) color space [16,21] to give $(L_1^*, a_1^*, b_1^*)$ and $(L_2^*, a_2^*, b_2^*)$ pair values using [Eqn. (8)-(11)]. $L_i^*$ correlates light-dark perception, $a_i^*$ correlates red-green chroma perception, and $b_i^*$ correlates blue-yellow chroma perception [16,21].

$$L_i^* = 116 f\left(\frac{Y_i}{Y_n}\right) - 16, \quad i = \{1,2\} \tag{8}$$

$$a_i^* = 500 \left[f\left(\frac{X_i}{X_n}\right) - f\left(\frac{Y_i}{Y_n}\right)\right], \quad i = \{1,2\} \tag{9}$$

$$b_i^* = 200 \left[f\left(\frac{Y_i}{Y_n}\right) - f\left(\frac{Z_i}{Z_n}\right)\right], \quad i = \{1,2\} \tag{10}$$

$$f(\omega) = \begin{cases} \omega^{\frac{1}{3}}, & \omega > \left(\frac{24}{116}\right)^3 \\ \left(\frac{841}{108}\right)\omega + \frac{16}{116}, & w \leq \left(\frac{24}{116}\right)^3 \end{cases} \tag{11}$$

In [Eqn. (8)-(10)], $(X_i, Y_i, Z_i)$ represent the tristimulus values of a color $i$, while $(X_n, Y_n, Z_n)$ represent the XYZ tristimulus values of an illuminant. In our case, the D65 illuminant takes values of $X_n = 95.04$, $Y_n = 100.00$, and $Z_n = 108.88$ [22].

*Step 2: Calculate Modified Chroma & Modified Hue Angle*

Chroma and hue angle are predictors derived from the cylindrical coordinate representation of the 1976 CIE (L*a*b*) color space [21]. However, modified derivations of chroma and hue angle need to be made because $a_i^*$ [Eqn. (9)] does not retain a uniform color space for neutral colors [19]. Moving forward, a modified value will be represented by the prime ($'$) instead of the asterisk ($*$). $a_i'$, the modified $a_i^*$, will have increased magnitudes for colors at low chroma and approximate magnitudes to $a_i^*$ for colors at high chroma [Eqn. (12)] [19]. No modifications need to be made to $L_i^*$ or $b_i^*$ [19]; however, the prime representations $L_i'$ [Eqn. (13)] and $b_i'$ [Eqn. (14)] will be used for the sake of uniformity.

$$a_i' = a_i^*(1 + G), \quad i = \{1,2\} \tag{12}$$

$$L_i' = L_i^*, \quad i = \{1,2\} \tag{13}$$

$$b_i' = b_i^*, \quad i = \{1,2\} \tag{14}$$

where

$$G = 0.5 \left[1 - \left(\frac{\overline{C_{ab}^*}^7}{\overline{C_{ab}^*}^7 + 25^7}\right)^{0.5}\right] \tag{15}$$

where $\overline{C_{ab}^*}$ [Eqn. (16)] is the arithmetic mean of the chroma $C_{ab}^*$ values [Eqn. (17)] for a pair of samples [16,19,21].

$$\overline{C_{ab}^*} = \frac{C_{ab_1}^* + C_{ab_2}^*}{2} \tag{16}$$

$$C_{ab_i}^* = \left(a_i^{*2} + b_i^{*2}\right)^{0.5}, \quad i = \{1,2\} \tag{17}$$

With $a'_i$ calculated, it becomes possible to calculate the modified chroma $C'_i$ [Eqn. (18)] and the modified hue angle $h'_i$ [Eqn. (19)].

$$C'_i = \left(a'^2_i + b'^2_i\right)^{0.5}, \qquad i = \{1,2\} \tag{18}$$

$$h'_i = \tan^{-1}\left(\frac{b'_i}{a'_i}\right), \qquad i = \{1,2\} \tag{19}$$

*Step 3: Calculate Differences in Lightness, Chroma, and Hue*

The difference in lightness [Eqn. (20)], chroma [Eqn. (21)], and hue [Eqn. (22)] between two color samples are next calculated using the modified values derived in the previous step.

$$\Delta L' = L'_1 - L'_2 \tag{20}$$

$$\Delta C' = C'_1 - C'_2 \tag{21}$$

$$\Delta H' = 2(C'_1 C'_2)^{0.5} \sin\left(\frac{\Delta h'}{2}\right) \tag{22}$$

where

$$\Delta h' = \begin{cases} 0, & C'_1 C'_2 = 0 \\ h'_1 - h'_2, & C'_1 C'_2 \neq 0; |h'_1 - h'_2| \leq 180° \\ (h'_1 - h'_2) - 360, & C'_1 C'_2 \neq 0; (h'_1 - h'_2) > 180° \\ (h'_1 - h'_2) + 360, & C'_1 C'_2 \neq 0; (h'_1 - h'_2) < -180° \end{cases} \tag{23}$$

The difference in the modified hue angle $\Delta h'$ in [Eqn. (22)] uses a four-quadrant arctangent. When evaluating the mean difference in the modified hue angle $\frac{\Delta h'}{2}$, a discontinuity emerges where the absolute difference in the modified hue angles is greater than 180°, as illustrated in Ref. [20]. In this case, the two colors are not in the same quadrant. To account for this, conditional statements [Eqn. (23)] are needed [19,20].

*Step 4: Calculate Parametric, Weighting, and Rotation Functions*

In addition to the terms calculated in the prior subsections, the CIEDE2000 equation [Eqn. (7)] contains parametric factors $k_L$, $k_C$, and $k_H$ [Eqn. (24)] to account for variations in overall perceptual sensitivity under a given set of experimental conditions [19]. Under the reference conditions outlined in Ref. [19], the parametric factors will all equal one.

$$k_L = k_C = k_H = 1 \tag{24}$$

The CIEDE2000 [Eqn. (7)] equation also contains weighting functions $S_L$ [Eqn. (25)], $S_C$ [Eqn. (26)], and $S_H$ [Eqn. (27)] to adjust the total equation for variation in perceived color-difference magnitude for lightness, chroma, or hue, respectively, with variation in the location of the color-difference pair in the $L'_i$, $a'_i$, $b'_i$ coordinates [19].

$$S_L = 1 + \frac{0.015\,(\bar{L}' - 50)^2}{[20 + (\bar{L}' - 50)^2]^{0.5}} \tag{25}$$

$$S_C = 1 + 0.045\overline{C'} \tag{26}$$

$$S_H = 1 + 0.015\overline{C'}T \tag{27}$$

where $\overline{L'}$ [Eqn. (28)] is the arithmetic mean of the lightness between the pair of color samples [19], $\overline{C'}$ [Eqn. (29)] is the arithmetic mean of the modified chroma between the pair of color samples [19], and $T$ [Eqn. (30)] is a contraction of the equation to make it more readable.

$$\overline{L'} = \frac{L'_1 + L'_2}{2} \tag{28}$$

$$\overline{C'} = \frac{C'_1 + C'_2}{2} \tag{29}$$

$$T = 1 - 0.17\cos(\overline{h'} - 30) + 0.24\cos(2\overline{h'})$$
$$+ 0.32\cos(3\overline{h'} + 6) - 0.20\cos(4\overline{h'} - 63) \tag{30}$$

where $\overline{h'}$ [Eqn. (31)] is the arithmetic mean of the modified hue angle between the pair of color samples [19].

$$\overline{h'} = \frac{h'_1 + h'_2}{2} \tag{31}$$

Finally, the CIEDE2000 [Eqn. (7)] equation contains a rotation function $R_T$ [Eqn. (32)] [19]. The rotation function takes into account the interaction between chroma difference and hue difference in the blue region of the 1976 CIE (L*a*b*) color space [16,19,21].

$$R_T = -\sin(2\Delta\Theta)\,R_C \tag{32}$$

where

$$\Delta\Theta = 30\exp\left[-\left(\frac{\overline{h'} - 275}{25}\right)^2\right] \tag{33}$$

$$R_C = 2\left(\frac{\overline{C'}^7}{\overline{C'}^7 + 25^7}\right)^{0.5} \tag{34}$$

*Step 5: Calculate CIEDE2000*

All arguments needed to calculate the CIEDE2000 color difference [Eqn. (7)] have been defined in the previous two sub-sections (Step 3 and Step 4). The CIEDE2000 equation is pasted below for convenience:

$$\text{CIEDE2000} = \left[\left(\frac{\Delta L'}{k_L S_L}\right)^2 + \left(\frac{\Delta C'}{k_C S_C}\right)^2 + \left(\frac{\Delta H'}{k_H S_H}\right)^2 + R_T\left(\frac{\Delta C'}{k_C S_C}\right)\left(\frac{\Delta H'}{k_H S_H}\right)\right]^{0.5} \tag{7}$$

**S5. Results**

*S5.1. Geometric Parameters & Spectra Summaries for All Tested Nanoparticles*

The geometric parameters for tested nanoparticles are given in Table S3, organized by the amount of corner cut applied to the nanoparticles. Following this, Table S4 then provides a results summary of the reflectance spectra for all the tested nanoparticles, organized in the same way as Table S3. The full spectral data for each tested nanoparticle is provided as Dataset 1 in Ref. [23].

**Table S3. Geometric parameters for all tested nanoparticles.**

| Aspect Ratio | Nanoparticle bound length (nm) | Nanoparticle bound width (nm) | Nanoparticle height (nm) | Cut corner length (nm) | Cut corner width (nm) | Nanostructure Volume ($\times 10^7$ nm$^3$) |
|---|---|---|---|---|---|---|
| | | | Uncut | | | |
| 1:1.00 | 240.42 | 240.42 | 270 | 0 | 0 | 1.56 |
| 1:1.25 | 215.03 | 268.79 | 270 | 0 | 0 | 1.56 |
| 1:1.50 | 196.30 | 294.45 | 270 | 0 | 0 | 1.56 |
| 1:1.75 | 181.74 | 318.04 | 270 | 0 | 0 | 1.56 |
| 1:2.00 | 170.00 | 340.00 | 270 | 0 | 0 | 1.56 |
| | | | One-Quarter Cut | | | |
| 1:1.00 | 240.42 | 240.42 | 270 | 60.104 | 60.104 | 1.46 |
| 1:1.25 | 215.03 | 268.79 | 270 | 53.759 | 67.198 | 1.46 |
| 1:1.50 | 196.30 | 294.45 | 270 | 49.075 | 73.162 | 1.46 |
| 1:1.75 | 181.74 | 318.04 | 270 | 45.434 | 79.518 | 1.46 |
| 1:2.00 | 170.00 | 340.00 | 270 | 42.500 | 85.000 | 1.46 |
| | | | Two-Quarter Cut | | | |
| 1:1.00 | 240.42 | 240.42 | 270 | 120.21 | 120.21 | 1.17 |
| 1:1.25 | 215.03 | 268.79 | 270 | 107.52 | 134.40 | 1.17 |
| 1:1.50 | 196.30 | 294.45 | 270 | 98.158 | 147.22 | 1.17 |
| 1:1.75 | 181.74 | 318.04 | 270 | 90.869 | 159.02 | 1.17 |
| 1:2.00 | 170.00 | 340.00 | 270 | 85.000 | 170.000 | 1.17 |
| | | | Three-Quarter Cut | | | |
| 1:1.00 | 240.42 | 240.42 | 270 | 180.31 | 180.31 | 0.68 |
| 1:1.25 | 215.03 | 268.79 | 270 | 161.28 | 201.67 | 0.68 |
| 1:1.50 | 196.30 | 294.45 | 270 | 147.22 | 220.84 | 0.68 |
| 1:1.75 | 181.74 | 318.04 | 270 | 136.30 | 238.53 | 0.68 |
| 1:2.00 | 170.00 | 340.00 | 270 | 127.50 | 255.00 | 0.68 |

**Table S4. Summary of spectra results for all tested nanoparticles.**

| Aspect Ratio | Polarization (CW or CCW) | Resonant Wavelengths (nm) | Amplitude at Resonant Wavelengths | Full Width at Half Maximum (nm) | XYZ Tristimulus Values | Color (Hex) |
|---|---|---|---|---|---|---|
| | | | Uncut | | | |
| 1:1.00 | Clockwise | 550 | 1.00 | 10.4 | (13.6, 20.8, 4.2) | 598C18 |
| 1:1.00 | Counterclockwise | 550 | 1.00 | 10.4 | (13.6, 20.8, 4.2) | 598C18 |
| 1:1.25 | Clockwise | 516, 587 | 0.43, 0.53 | 10.9, 8.7 | (11.8, 14.9, 5.0) | 647230 |
| 1:1.25 | Counterclockwise | 516, 587 | 0.43, 0.53 | 10.9, 8.7 | (11.8, 14.9, 5.0) | 647230 |
| 1:1.50 | Clockwise | 486, 621 | 0.43, 0.53 | 15.5, 6.6 | (8.7, 8.7, 8.0) | 5C514B |
| 1:1.50 | Counterclockwise | 486, 621 | 0.23, 0.52 | 15.5, 6.6 | (8.7, 8.7, 8.0) | 5C514B |
| 1:1.75 | Clockwise | 464, 651 | 0.23, 0.52 | 10.1, 4.8 | (7.1, 6.5, 13.0) | 484564 |
| 1:1.75 | Counterclockwise | 464, 651 | 0.34, 0.51 | 10.1, 4.8 | (7.1, 6.5, 13.0) | 484564 |
| 1:2.00 | Clockwise | 446, 681 | 0.46, 0.51 | 6.7, 3.0 | (6.7, 5.7, 12.9) | 483D64 |
| 1:2.00 | Counterclockwise | 446, 681 | 0.46, 0.51 | 6.7, 3.0 | (6.7, 5.7, 12.9) | 483D64 |
| **One-Quarter Cut** | | | | | | |
| 1:1.00 | Clockwise | 536, 550 | 0.56, 0.61 | 4.6, 23.7 | (11.1, 18.6, 3.8) | 428716 |
| 1:1.00 | Counterclockwise | 536, 550 | 0.56, 0.61 | 4.6, 23.7 | (11.1, 18.6, 3.8) | 428716 |
| 1:1.25 | Clockwise | 508, 580 | 0.23, 0.51 | 13.0, 7.7 | (10.0, 12.0, 4.4) | 60652E |
| 1:1.25 | Counterclockwise | 508, 580 | 0.36, 0.54 | 12.1, 7.6 | (10.2, 13.3, 5.0) | 5B6C32 |
| 1:1.50 | Clockwise | 478, 614 | 0.14, 0.50 | 19.4, 5.0 | (7.6, 7.4, 6.6) | 594A44 |
| 1:1.50 | Counterclockwise | 478, 614 | 0.23, 0.53 | 16.0, 5.0 | (8.0, 7.7, 8.3) | 584B4E |
| 1:1.75 | Clockwise | 458, 646 | 0.27, 0.50 | 9.8, 2.4 | (5.9, 5.6, 10.2) | 424059 |
| 1:1.75 | Counterclockwise | 458, 646 | 0.36, 0.53 | 9.5, 2.4 | (6.3, 5.7, 12.4) | 433F62 |
| 1:2.00 | Clockwise | 441, 679 | 0.42, 0.17 | 5.4, 3.0 | (5.6, 5.1, 9.7) | 433C57 |
| 1:2.00 | Counterclockwise | 441, 679 | 0.50, 0.17 | 5.4, 2.9 | (5.8, 5.1, 11.0) | 433A5D |
| **Two-Quarter Cut** | | | | | | |
| 1:1.00 | Clockwise | 544 | 0.52 | 6.3 | (4.9, 8.8, 2.3) | 1C6117 |
| 1:1.00 | Counterclockwise | 544 | 0.52 | 6.3 | (4.9, 8.8, 2.3) | 1C6117 |
| 1:1.25 | Clockwise | 570 | 0.20 | 1.5 | (3.4, 4.1, 2.0) | 363C20 |
| 1:1.25 | Counterclockwise | 489, 570 | 0.12, 0.24 | 22.6, 1.5 | (3.7, 4.6, 3.2) | 32402D |
| 1:1.50 | Clockwise | - | - | - | (2.9, 3.5, 2.2) | 2F3824 |
| 1:1.50 | Counterclockwise | 460 | 0.14 | 6.1 | (3.4, 3.7, 5.1) | 2E373E |
| 1:1.75 | Clockwise | 439 | 0.05 | 4.5 | (3.0, 3.4, 2.7) | 31352A |
| 1:1.75 | Counterclockwise | 440 | 0.32 | 4.9 | (3.6, 3.4, 6.2) | 343146 |
| 1:2.00 | Clockwise | 425, 431 | 0.16 | 3.6 | (3.1, 3.3, 3.1) | 35322F |
| 1:2.00 | Counterclockwise | 425, 431 | 0.47, 0.16 | 3.8, 0.95 | (3.5, 3.2, 5.0) | 37303E |
| **Three-Quarter Cut** | | | | | | |
| 1:1.00 | Clockwise | - | - | - | (2.8, 2.3, 1.1) | 3E2416 |
| 1:1.00 | Counterclockwise | - | - | - | (2.8, 2.3, 1.1) | 3E2416 |
| 1:1.25 | Clockwise | - | - | - | (2.7, 2.3, 1.1) | 3D2416 |
| 1:1.25 | Counterclockwise | - | - | - | (2.7, 2.3, 0.89) | 3D2412 |
| 1:1.50 | Clockwise | - | - | - | (2.6, 2.3, 0.89) | 3B2612 |
| 1:1.50 | Counterclockwise | - | - | - | (2.6, 2.3, 0.82) | 332910 |
| 1:1.75 | Clockwise | - | - | - | (2.6, 2.4, 0.81) | 3B2610 |
| 1:1.75 | Counterclockwise | - | - | - | (2.6, 2.4, 0.81) | 3B2610 |
| 1:2.00 | Clockwise | - | - | - | (2.6, 2.4, 0.74) | 3B270E |
| 1:2.00 | Counterclockwise | - | - | - | (2.6, 2.4, 0.73) | 3B270E |

*S5.2. Three-Quarter Cut Structures*

Compared to the uncut, one-quarter cut, and two-quarter cut structures, the three-quarter cut structures fail to exhibit a resonance in reflectance (Fig. S17). Consequently, follow-up results and analyses for the three-quarter cut structures were mostly omitted from the Main Text. Three-quarter cut structures have the smallest amount of volume, and we suspect this reduction in volume removes the nanoparticles' ability to support resonance. We made a similar hypothesis in our previous work [24].

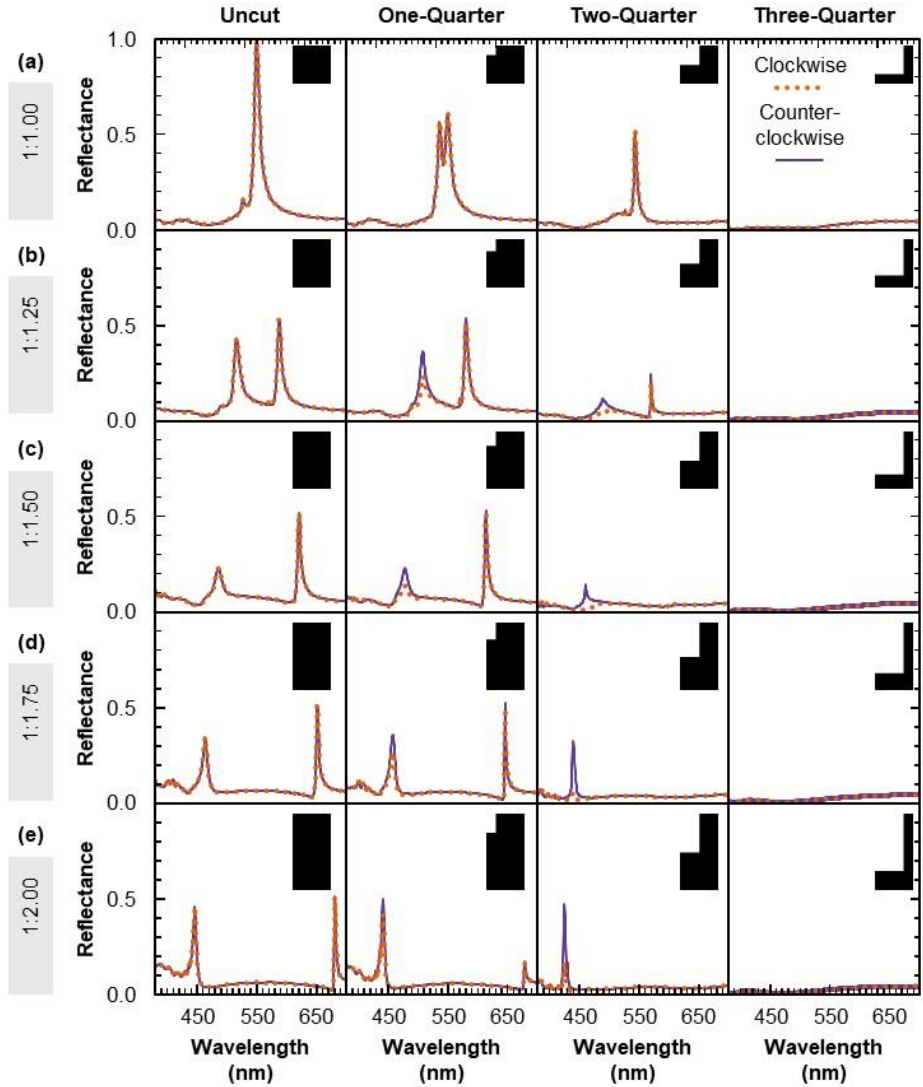

Fig. S17. Reflectance plots for the uncut (first column), one-quarter cut (second column), two-quarter cut (third column), and three-quarter cut (last column) for the (a) 1:1.00, (b) 1:1.25, (c) 1:1.50, (d) 1:1.75, and (e) 1:2.00 structures illuminated with clockwise (dotted orange line) or counterclockwise (solid purple line) CPL. The three-quarter cut structures (last column) did not show any resonance with either CPL orientation.

*S5.3. Full Domain Net Electric Flux Results*

Following the observed far-field spectra of the two-quarter cut structures in Fig. S16, we analyzed the net electric flux for the x-, y-, and z-components of the 1:1.00 (Fig. S18), 1:1.25 (Fig. S19), 1:1.50 (Fig. S20), 1:1.75 (Fig. S21), and 1:2.00 (Fig. S22) structures and their resonance-bearing corner cuts. This data is provided as Dataset 1 in Ref. [23]. We did notice that the corner cuts, in all cases, caused a deviation in the net electric flux between the clockwise and counterclockwise CPL illuminations, which we've attributed to disruptions of the mirror symmetry of the structures in the Main Text of this work.

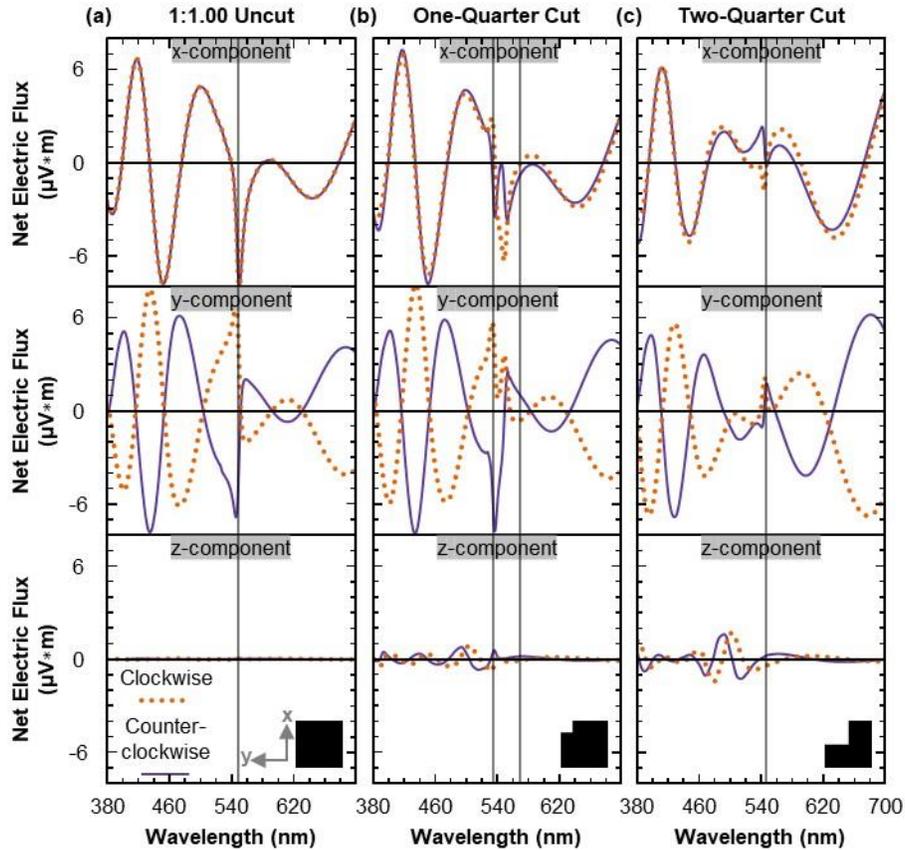

Fig. S18. (a-c) Net electric flux graphs of the 1:1.00 (a) uncut, (b) one-quarter cut, and (c) three-quarter cut structures. (Top) x-component of the net electric flux; (middle) y-component of the net electric flux; (bottom) z-component of the net electric flux. The x- and y-components are aligned with the in-plane scattering. The z-component is aligned with the out-of-plane scattering. All graphs plot net electric fluxes resultant from exciting the nanostructure arrays with clockwise (orange dotted line) or counterclockwise (purple line) CPL. Solid, faint gray lines descending through the plots point to the respective resonance wavelengths on the horizontal axes.

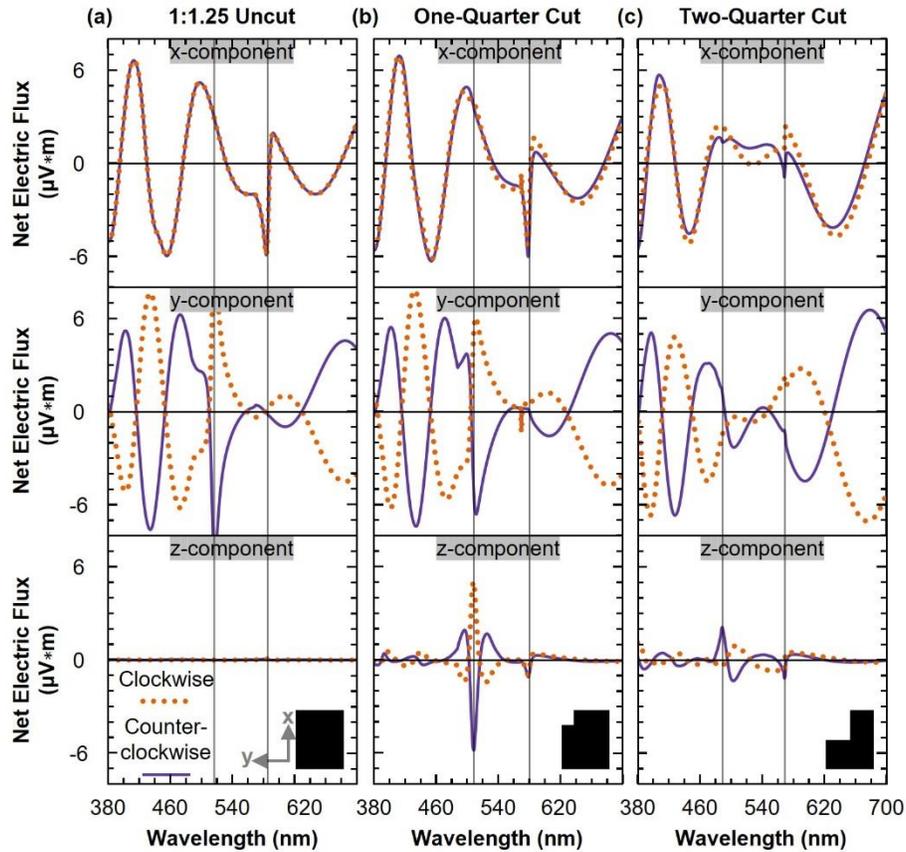

Fig. S19. (a-c) Net electric flux graphs of the 1:1.25 (a) uncut, (b) one-quarter cut, and (c) three-quarter cut structures. (Top) x-component of the net electric flux; (middle) y-component of the net electric flux; (bottom) z-component of the net electric flux. The x- and y-components are aligned with the in-plane scattering. The z-component is aligned with the out-of-plane scattering. All graphs plot net electric fluxes resultant from exciting the nanostructure arrays with clockwise (orange dotted line) or counterclockwise (purple line) CPL. Solid, faint gray lines descending through the plots point to the respective resonance wavelengths on the horizontal axes.

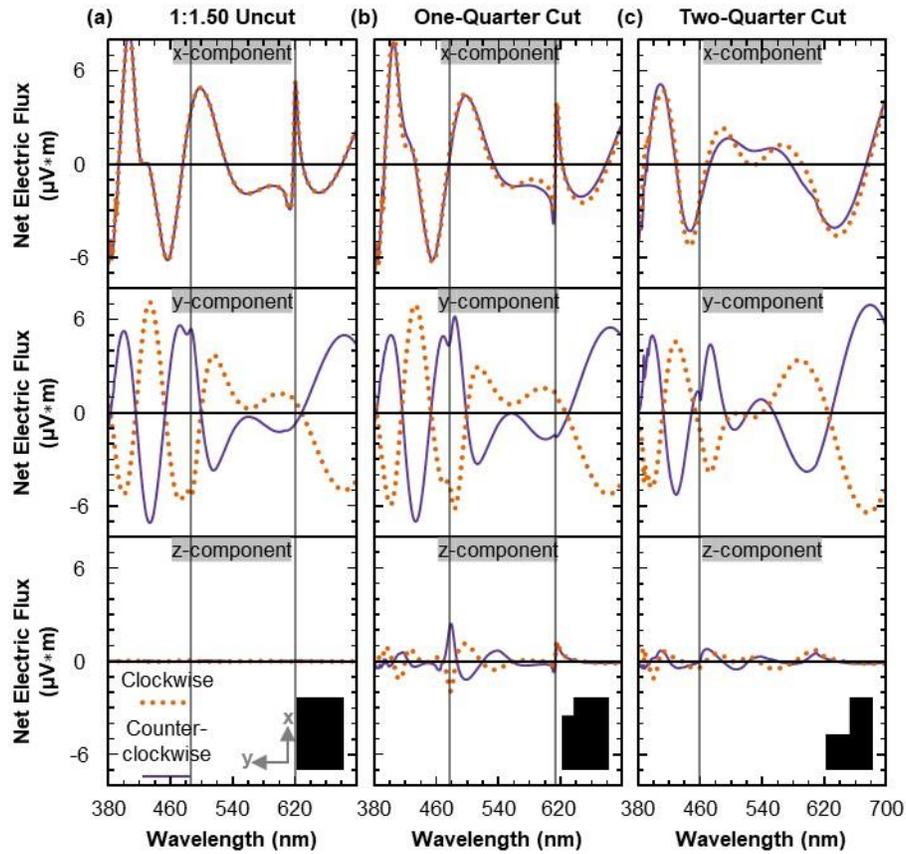

Fig. S20. (a-c) Net electric flux graphs of the 1:1.50 (a) uncut, (b) one-quarter cut, and (c) three-quarter cut structures. (Top) x-component of the net electric flux; (middle) y-component of the net electric flux; (bottom) z-component of the net electric flux. The x- and y-components are aligned with the in-plane scattering. The z-component is aligned with the out-of-plane scattering. All graphs plot net electric fluxes resultant from exciting nanostructure arrays with clockwise (orange dotted line) or counterclockwise (purple line) CPL. Solid, faint gray lines descending through the plots point to the respective resonance wavelengths on the horizontal axes.

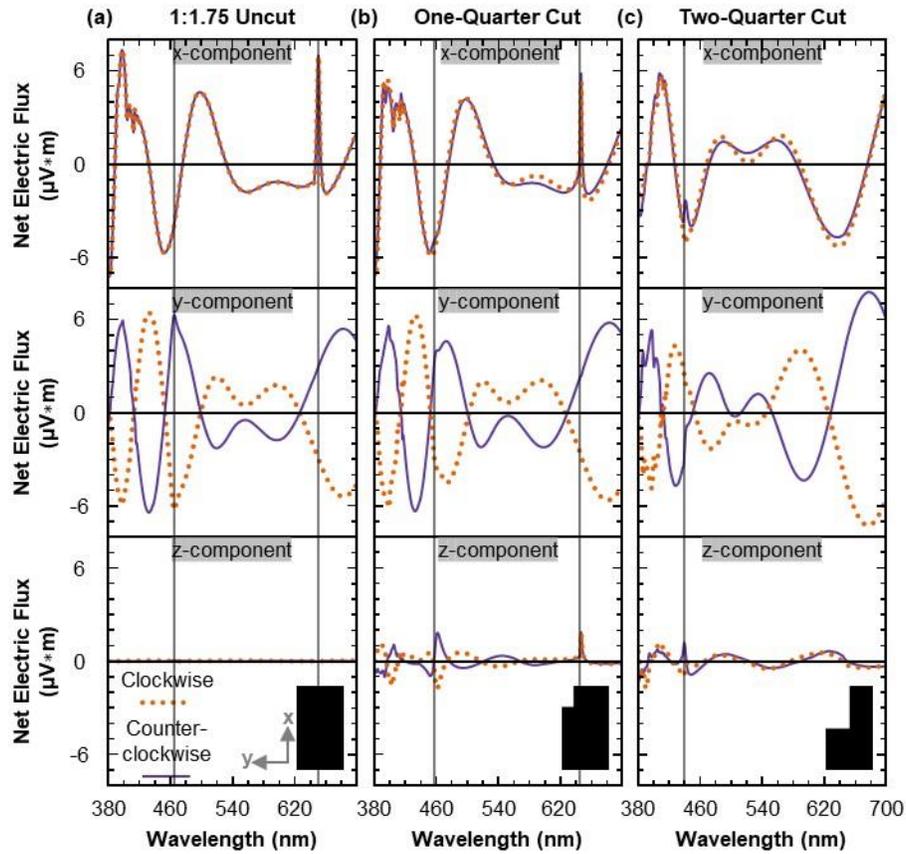

Fig. S21. (a-c) Net electric flux graphs of the 1:1.75 (a) uncut, (b) one-quarter cut, and (c) three-quarter cut structures. (Top) x-component of the net electric flux; (middle) y-component of the net electric flux; (bottom) z-component of the net electric flux. The x- and y-components are aligned with the in-plane scattering. The z-component is aligned with the out-of-plane scattering. All graphs plot net electric fluxes resultant from exciting nanostructure arrays with clockwise (orange dotted line) or counterclockwise (purple line) CPL. Solid, faint gray lines descending through the plots point to the respective resonance wavelengths on the horizontal axes.

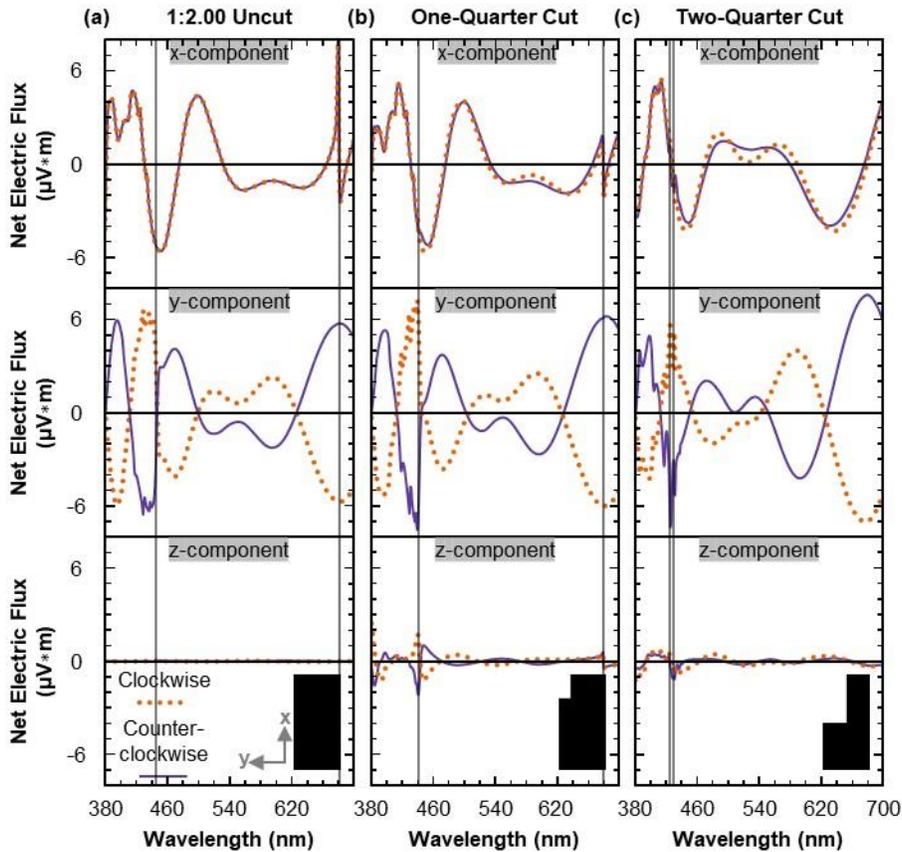

Fig. S22. (a-c) Net electric flux graphs of the 1:2.00 (a) uncut, (b) one-quarter cut, and (c) three-quarter cut structures. (Top) x-component of the net electric flux; (middle) y-component of the net electric flux; (bottom) z-component of the net electric flux. The x- and y-components are aligned with the in-plane scattering. The z-component is aligned with the out-of-plane scattering. All graphs plot net electric fluxes resultant from exciting nanostructure arrays with clockwise (orange dotted line) or counterclockwise (purple line) CPL. Solid, faint gray lines descending through the plots point to the respective resonance wavelengths on the horizontal axes.

### S5.4. Colorimetry Results

Fig. 5 of the Main Text revealed that the two-quarter cut structures showed the largest color-difference score between exciting a nanostructure array with clockwise or counterclockwise illuminations. Fig. S23 shows the CIE 1931 2-degree Standard Observer color space plots for all our structures under both CPL polarizations. Visually, we notice that there is a greater increase in the distance between the plotted color coordinates for the two-quarter cut structures.

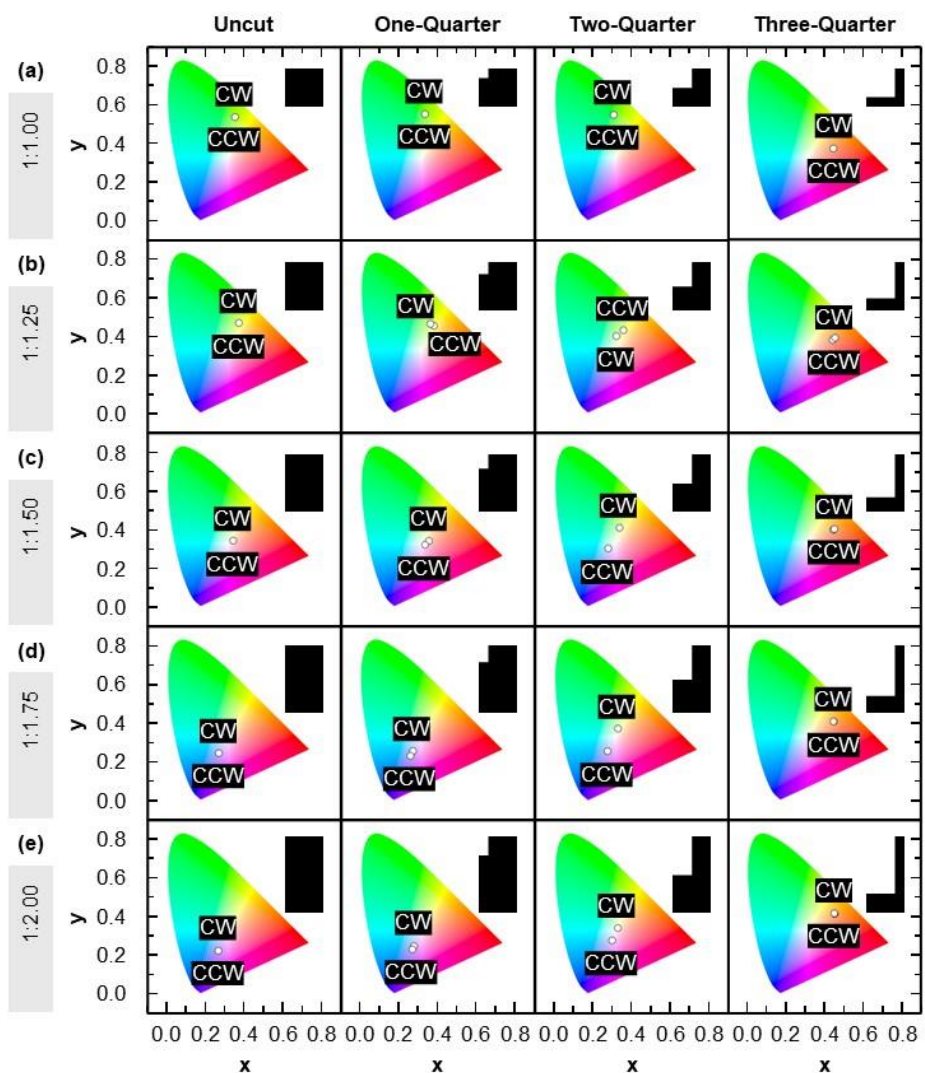

Fig. S23. CIE 1931 2-degree Standard Observer color space plots for the uncut (first column), one-quarter cut (second column), two-quarter cut (third column), and three-quarter cut (last column) for the (a) 1:1.00, (b) 1:1.25, (c) 1:1.50, (d) 1:1.75, and (e) 1:2.00 structures illuminated with clockwise (labelled "CW") or counterclockwise (labelled "CCW") CPL. The three-quarter cut structures (last column) did not show any resonance with either CPL orientation.

### S5.5. Colorimetry Descriptive Statistics

Table S8 provides data and descriptive statistics of the color-distance results for all tested nanoparticles. The data in Table S5 is also portrayed graphically in Fig. 5 of the Main Text. Of note, the two-quarter cut structures show a greater mean and median scores than the entire mean and median scores of the entire dataset, even when individually compared to the other corner cuts.

Table S8. Colorimetry descriptive statistics.

| **CIEDE2000 Color-Difference Scores** |
|---|

| Aspect Ratio | Uncut | One-Quarter Cut | Two-Quarter Cut | Three-Quarter Cut |
|---|---|---|---|---|
| **1:1.00** | 0.00 | 0.00 | 0.00 | 0.00 |
| **1:1.25** | 0.00 | 5.82 | 4.55 | 2.53 |
| **1:1.50** | 0.00 | 8.17 | 16.47 | 0.68 |
| **1:1.75** | 0.00 | 4.67 | 24.65 | 0.25 |
| **1:2.00** | 0.00 | 2.78 | 15.54 | 0.80 |
| **Descriptive Statistics with All Data** | | | | |
| **Mean** | 0.00 | 4.29 | 12.24 | 0.85 |
| **Median** | 0.00 | 4.67 | 15.54 | 0.68 |
| **Standard Deviation** | 0.00 | 3.09 | 9.90 | 0.99 |
| **Standard Error** | 0.00 | 1.38 | 4.43 | 0.44 |
| **Variance** | 0.00 | 9.56 | 97.94 | 0.98 |
| **Descriptive Statistics Without 1:1.00 Data** | | | | |
| **Mean** | 0.00 | 5.36 | 15.30 | 1.07 |
| **Median** | 0.00 | 5.25 | 16.01 | 0.74 |
| **Standard Deviation** | 0.00 | 2.25 | 8.25 | 1.00 |
| **Standard Error** | 0.00 | 1.13 | 4.13 | 0.50 |
| **Variance** | 0.00 | 5.08 | 68.14 | 1.01 |
| **Descriptive Statistics of Entire Dataset** | | | | |
| **Mean** | 4.35 | | | |
| **Median** | 0.74 | | | |
| **Standard Deviation** | 6.71 | | | |
| **Standard Error** | 1.50 | | | |
| **Variance** | 45.06 | | | |